\documentclass[floatfix,groupedaddress,twocolumn,showpacs,amsmath,pra,aps,amssymb,longbibliography,nofootinbib]{revtex4-1}

\usepackage{graphicx}
\usepackage{xcolor}
\usepackage{lipsum}
\usepackage{overpic}
\usepackage[english]{babel}
\usepackage{amsmath,amsfonts,amssymb,float}
\usepackage[titletoc,title]{appendix} 
\usepackage{soul}
\usepackage{mathtools}
\usepackage{bbold}

\usepackage[unicode]{hyperref}
\hypersetup{
	unicode=true, % nonLatin characters in Acrobat's bookmarks
	a4paper=true,
	plainpages=false,
	colorlinks=true,% false: boxed links; true: colored links
	linkcolor=blue,% color of internal links
	citecolor=blue,% color of links to bibliography
	filecolor=blue,% color of file links
	urlcolor=blue% color of external links
}
\newcommand\widthscale{1.0}

\newcommand{\beq}{\begin{equation}}
\newcommand{\eeq}{\end{equation}}
\newcommand{\bea}{\begin{eqnarray}}
\newcommand{\eea}{\end{eqnarray}}
\newcommand{\BEQAL}{\begin{align}}
\newcommand{\EEQAL}{\end{align}}

\newcommand{\comment}[1]{{}}

\newcommand{\commentout}[1]{{}}

\newcommand{\ket}[1]{\ensuremath{\left\vert #1 \right\rangle}}
\begin{document}

\title{Bistable optical transmission through arrays of atoms in free space}
\author{C. D. Parmee}
\affiliation{Department of Physics, Lancaster University, Lancaster, LA1 4YB, United Kingdom}
\author{J. Ruostekoski}
\affiliation{Department of Physics, Lancaster University, Lancaster, LA1 4YB, United Kingdom}
\date{\today}

\begin{abstract}
We determine the transmission of light through a planar atomic array beyond the limit of low light intensity that displays optical bistability in the mean-field regime.
We develop a theory describing the intrinsic optical bistability, which is supported purely by resonant dipole-dipole interactions in free space, showing how bistable light amplitudes 
exhibit both strong cooperative and weak single-atom responses and how they depend on the underlying low light intensity collective excitation eigenmodes.
Similarities of the theory with optical bistability in cavities are highlighted, while recurrent light scattering between atoms takes on the role of cavity mirrors.
Our numerics and analytic estimates show a sharp variation in the extinction, reflectivity, and group delays of the array, with the incident light completely extinguished 
up to a critical intensity well beyond the low light intensity limit.
Our analysis paves a way for collective nonlinear optics with cooperatively responding dense atomic ensembles.
\end{abstract}

\maketitle

\section{Introduction}

Optical bistability is a intriguing nonlinear phenomena for atoms, where two possible states of a system coexist for the same parameters.
It has been studied theoretically for a long time in the regime where atoms couple to a single cavity mode and atomic positions and spatial dependence of interactions do not play a role~\cite{Lugiato1984,bonifacio1976,Bonifacio1978,Carmichael1977,Agrawal79,Drummond1980,Drummond1981,Savage1988a,Carmichael1986a}.
It has also been observed experimentally~\cite{Gibbs1976,Rosenberger1983,Orozco1984,Rempe1991,Gothe2019,Lambrecht1995}, sometimes alongside a rich phenomenology of other optically-induced phases~\cite{Caballero-Benitez2015,Landig2016,Ivanov2020,Muniz2020}.
However, bistability can also be found less commonly in systems where it is intrinsic, generated only by interactions within the sample, and has so far been observed in Yb$^{3+}$ ions in solid-state crystals at cryogenic temperatures~\cite{Hehlen1994} and in highly-excited Rydberg atoms in the microwave regime~\cite{Carr2013}. Intrinsic bistability was thought to be unachievable for atoms in the optical regime, but recent theoretical studies of many-body systems suggest that interaction-mediated bistability is more generic and possible in a variety of systems with short- and long-range interactions~\cite{Lee2011,Sibalic2016,Parmee2018}.
In particular, we recently demonstrated~\cite{parmee2020} that intrinsic bistability and optically-induced phases emerge in arrays of atoms at sufficiently high densities, due to resonant light-mediated dipole-dipole (DD) interactions and that these could be identified
in coherently and incoherently scattered light.

Several experiments on interactions of light with atomic ensembles have in recent years achieved such cold temperatures and high atom densities~\cite{Bender2010,BalikEtAl2013,Pellegrino2014a,KEM14,Jenkins_thermshift,Jennewein_trans,vdStraten16,Dalibard_slab,Machluf2018,Saint-Jalm2018} that the collective optical responses can start deviating~\cite{Javanainen2014a,Jenkins_thermshift} from those of thermal or low-density samples. 
The relevant density scale is the number of atoms per cubic wavenumber $k$ of the resonant light, which takes nonnegligible values also for atoms in optical lattices. 
In particular, a Mott-insulator state of ${}^{87}\text{Rb}$ atoms was now studied~\cite{Rui2020} in an optical lattice with near unit filling, where a subradiant eigenmode with a spatially uniform phase profile was driven by the incident field in the limit of low light intensity (LLI), and observed in a narrowed transmission resonance for light. 
Collective interaction of light with closely related arrays of atoms has attracted considerable theoretical interest~\cite{Clemens2003a,Antezza2009,Jenkins2012a,Olmos13,Castin13,Bettles2016,Facchinetti16,Yoo2016,Kramer2016,Jen17,Shahmoon,Sutherland1D,Perczel2017,Bettles2017,Facchinetti18,Bhatti18,Asenjo_prx,Zhang2018,Plankensteiner2017,Grankin18,Guimond2019,ballantine2020,Javanainen19,Henriet2018,Mkhitaryan18,Qu19,Needham19,Ritsch_subr,Asenjo-Garcia19,williamson2020b,Cidrim20,Ballantine20Huygens,Alaee20,Ballantine20Toroidal,Yoo20,parmee2020,Bettles2020,Shahmoon19,Orioli19,Zhang20}.

Here we analyze transmission of light through a dense planar array of cold atoms beyond the limit of LLI, when the atoms respond nonlinearly to light. 
We formulate a theory for optical bistability of free-space atomic arrays that in general depends on the underlying LLI mode, expanding our earlier analysis of Ref.~\cite{parmee2020}. 
In some cases, the theory can be solved even analytically, with sufficiently small atomic separations $(ka)<(\pi/3)^{1/2}$ needed for bistability of spatially uniform modes.
The bistability threshold  $ka\sim 1$ applies even for the case of just two atoms and equals the separation at which the single-atom linewidth $\gamma$ becomes less than
the collective line shift, originating from recurrent scattering  where the light is scattered more than once by the same atom~\cite{Ishimaru1978,Lagendijk,vantiggelen90,Morice1995a,Ruostekoski1997a,Kwong19}.

We find that the transmitted light exhibits a bistable solution of both ``cooperative'' and ``single-atom'' branches, which we obtain approximate analytic solutions for at high atomic densities. The cooperative branch represents a collective response where atoms strongly absorb the incident light, with high extinction and weak incoherent scattering, while the single-atom branch represents an independent response with atoms weakly absorbing the incident field, with low extinction and strong incoherent scattering.
In particular, the cooperative branch can completely extinguish the incident light up to a large critical intensity, $I_c/I_{\rm{sat}} \simeq 155$, well beyond the LLI limit. Beyond this intensity, a sharp change in the transmission behavior of the array occurs, as light begins to transmit through the lattice. 
By varying the frequency and intensity of the incident light, we find hysteresis between the branches can occur, which is observable by jumps in the extinction, reflectivity and phase shifts of the light.
We also find the loss of one of the branches at the edge of the bistability region is associated with a first order phase transition, resulting in a divergence of the group delay and critical slowing, where increasingly long times are needed to reach the steady state.

The emergence of optical bistability in a collectively responding system of regularly spaced array of atoms has a surprisingly close analogy with the optical bistability in cavities. The presence of closely spaced atoms and strong DD interactions modify the
effective Purcell factors of atoms, due to substantial recurrent light scattering. This is reminiscent of the effect of a cavity in which case a cooperative
response results from an atom repeatedly scattering the same photon that bounces between the cavity mirrors. The recurrent scattering in cavities is quantified by a cooperativity parameter $C = g^2/2\gamma\kappa$, which depends on the cavity  linewidth $\kappa$ and atom-cavity coupling $g$.
For specific parameter values the analogy between bistability in the two systems becomes direct, with the same equations governing the relationship between the incident and total light field, in which case we can define 
a cooperativity parameter for atomic arrays as $C=\tilde{\gamma}/2\gamma$, for the collective linewidth $\gamma+\tilde{\gamma}$. The bistability in both systems then emerges for a sufficiently strong cooperative response when $C>4$. 

The layout of this paper is as follows. In Sec.~\ref{model}, we present the model. In Sec.~\ref{2Darrays}, we present the theoretical description of optical bistability in 2D planar arrays, before studying transmission properties in Sec.~\ref{transmission}. Finally, in Sec.~\ref{discussion}, we discuss our results and draw conclusions. In Appendices A and B, we give further details on the comparison between bistability in cavities and atom arrays, and on the transmission results, respectively.

\section{Model}\label{model}

\subsection{Atoms and light fields}

We consider a system of $N$ two-level cold atoms interacting with light and trapped in a two-dimensional (2D) array with one atom per site. The electrodynamics are expressed in the {\it length-gauge}, obtained by the Power-Zienau-Woolley transformation~\cite{PowerZienauPTRS1959,Woolley1971a,CohenT}, where the basic dynamical variable for light is the electric displacement vector, $\hat{\textbf{D}}(\textbf{r})=\hat{\textbf{D}}{}^+(\textbf{r})+\hat{\textbf{D}}{}^-(\textbf{r})$. The positive frequency component is $\hat{\textbf{D}}{}^+(\textbf{r}) = \sum_q\zeta\hat{a}_qe^{\rm{i}\textbf{q}\cdot\textbf{r}}\hat{\textbf{e}}_q$, with $\hat{\textbf{D}}{}^-(\textbf{r})=[\hat{\textbf{D}}{}^+(\textbf{r})]{}^{\dagger}$, where $\zeta = \sqrt{\hbar\omega_q\epsilon_0/2V}$, and we have introduced the mode frequency, $\omega_q$, polarization $\hat{\textbf{e}}_q$, photon annihilation operator, $\hat{a}_q$ and mode volume, $V$.
The polarization of the atoms is expressed through the polarization vector with positive frequency component,
\begin{equation}
\hat{\textbf{P}}{}^+(\textbf{r})=\sum_l\delta(\textbf{r}-\textbf{r}_l)\textbf{d}_{ge} \hat{\sigma}_l^-.
\end{equation}
The atoms are at fixed lattice sites $\textbf{r}_l$, and we have introduced the dipole moment, $\textbf{d}_{ge}$, and the lowering operator $\hat{\sigma}_l^{-}=|g\rangle_{l}\mbox{}_{l}\langle e|=(\hat{\sigma}_l^+)^{\dagger}$, with $\ket{e}_l$ and $\ket{g}_l$ denoting the excited and ground state of the two-level atom on site $l$, respectively.
We assume the atoms are illuminated by a near monochromatic incident field, $\hat{\textbf{D}}{}^+_F(\textbf{r})=\epsilon_0\boldsymbol{\mathcal{E}}{}^+(\textbf{r})$ where $ \boldsymbol{\mathcal{E}}{}^+(\textbf{r})=\mathcal{E}_0\hat{\textbf{e}}e^{\text{i}\textbf{k}\cdot\textbf{r}}$, with wavevector $\textbf{k}$ and frequency $\omega = c|\textbf{k}|=ck$, and express observables in terms of slowly varying field amplitudes and atomic variables, $\hat{\textbf{D}}{}^+e^{i\omega t} \rightarrow \hat{\textbf{D}}{}^+$ and $\hat{\sigma}{}^{-}_le^{i\omega t} \rightarrow \hat{\sigma}{}^{-}_l$.
The incident light is expressed through the Rabi frequency, ${\cal R}_l=\textbf{d}_{ge}^*\cdot\boldsymbol{\mathcal{E}}{}^+(\textbf{r}_l)/\hbar$ acting on an atom at lattice site $l$, and the incident and saturation intensity,
\begin{equation}\label{intensity}
\begin{split}
&\frac{I_l}{I_{\rm sat}}=2\frac{|{\cal R}_l|^2}{\gamma^2},\quad I_{\rm sat}=\hbar c \frac{ 4\pi^2 \gamma}{3\lambda^3}.
\end{split}
\end{equation}
Throughout the paper, we assume $\textbf{d}_{ge}=\mathcal{D}\hat{\textbf{e}}$, where $\mathcal{D}$ is the reduced dipole matrix element.

Integrating over all space, the system Hamiltonian is~\cite{Ruostekoski1997a}
\begin{equation}\label{Eq:Hamiltonian}
\begin{split}
\hat{H}&=\sum_{q}\hbar\omega_{q}\hat{a}_q^{\dagger}\hat{a}_q+\sum_{l}\Delta_l\hat{\sigma}^{ee}_l\\
&+ \frac{1}{2\epsilon_0}\int\hat{\textbf{P}}(\textbf{r})\cdot \hat{\textbf{P}}(\textbf{r})d^3\textbf{r}-\frac{1}{\epsilon_0}\int\hat{\textbf{D}}(\textbf{r})\cdot \hat{\textbf{P}}(\textbf{r})d^3\textbf{r}.
\end{split}
\end{equation} 
The first term is the Hamiltonian of the free electromagnetic field. The second term is the laser frequency detuning from the atomic resonance, $\Delta_l=\omega-\omega^l_{eg}$, where $\omega^l_{eg}$ is the transition frequency of an atom on site $l$, 
and $\hat{\sigma}^{ee}_l=\hat{\sigma}_l^{+}\hat{\sigma}_l^{-}$.  
The third term is the self-polarization, which is zero for nonoverlapping point atoms. 
The final term is the interaction of the electric displacement field with the atomic polarization. After carrying out the spatial integration, we also make a rotating wave approximation in the Hamiltonian, Eq.~\eqref{Eq:Hamiltonian}, to remove the fast co-rotating terms.

The total electric field $\hat{\textbf{E}}{}^+(\textbf{r}) =  \boldsymbol{\mathcal{E}}{}^+(\textbf{r})+\hat{\textbf{E}}{}^+_s(\textbf{r})$ is obtained using $\hat{\textbf{D}}(\textbf{r}) = \epsilon_0\hat{\textbf{E}}(\textbf{r})+\hat{\textbf{P}}(\textbf{r})$ from the scattered field,
\begin{equation} \label{Efieldscattered}
\epsilon_0\hat{\textbf{E}}{}^+_{s}(\textbf{r}) =\sum_l\mathsf{G}(\textbf{r}-\textbf{r}_l)\textbf{d}_{ge} \hat{\sigma}_l^-,
\end{equation}
where the field satisfies Maxwell's wave equation with an atomic polarization source~\cite{Ruostekoski1997a} and the dipole radiation kernel acting on a dipole located at the origin, with $r= |\textbf{r}|$ and $\hat{\mathbf{r}}=\textbf{r}/r$, is given by the familiar dipole radiation pattern~\cite{Jackson,BOR99}
\begin{align}\label{Gdef}
\mathsf{G}(\mathbf{r})\mathbf{d}&=-\frac{\mathbf{d}\delta(\mathbf{r})}{3}+\frac{k^3}{4\pi}\Bigg\{\left(\hat{\mathbf{r}}\times\mathbf{d}\right)\times\hat{\mathbf{r}}\frac{e^{ikr}}{kr}\nonumber\\
&\phantom{==}-\left[3\hat{\mathbf{r}}\left(\hat{\mathbf{r}}\cdot\mathbf{d}\right)-\mathbf{d}\right]\left[\frac{i}{(kr)^2}-\frac{1}{(kr)^3}\right]e^{ikr}\Bigg\}.
\end{align}

\begin{figure}
	\hspace*{0cm}
	\includegraphics[width=0.9\columnwidth]{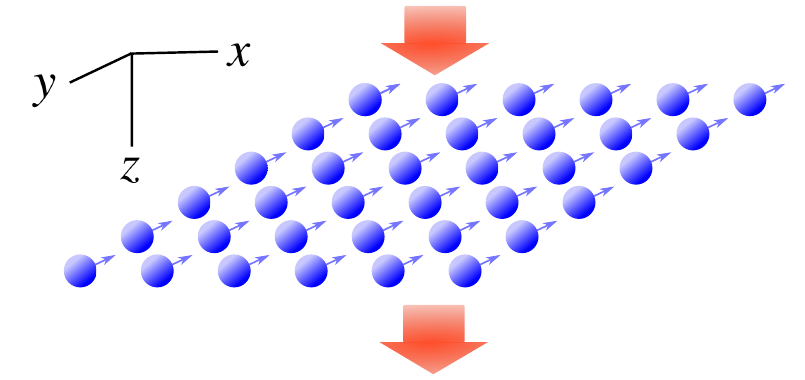}
	\vspace{-0.2cm}
	\caption{Transmission through an array of atoms. An incident field drives a large collective response at low intensities and a weak single-atom response at high intensities. For intermediate intensities, bistability between the two responses is possible.}
	\label{Fig:Model}
\end{figure}

\subsection{Mean-field approximation}

\subsubsection{Mean-field equations}
We now describe the dynamical evolution of the atomic coherences and excited level population in the mean-field regime where quantum fluctuations between different atoms~\cite{Bettles2020} are ignored. 
The dynamics of the system are obtained by solving the Heisenberg equations of motion for the atomic operators $\hat{\sigma}_l^{ee}$ and $\hat{\sigma}_l^{-}$, assuming a Born-Markov approximation to eliminate the electric field operators, $\hat{a}_q$. A Gutzwiller mean-field approximation is then implemented, corresponding to the factorization of internal level correlations,
\begin{equation}\label{Factorisation}
\begin{split}
\langle \hat{\sigma}{}_i^{\alpha}\hat{\sigma}{}_j^{\beta}\rangle\approx \langle \hat{\sigma}{}_i^{\alpha}\rangle\langle \hat{\sigma}{}_j^{\beta}\rangle, \quad i \neq j
\end{split}
\end{equation}
where, because atoms are at fixed positions with no position fluctuations, there are no light-induced correlations~\cite{Lee16,Bettles2020} between the atoms after the factorization. 
The system dynamics are then described by the following nonlinear equations
\begin{subequations}\label{SpinEqs}
	\begin{align}
	\begin{split}
	\dot{\rho}^{(l)}_{ge}=&\left(\text{i}\Delta_l-\gamma\right) {\rho}^{(l)}_{ge}\\
	&-\text{i}(2{\rho}^{(l)}_{ee}-1)\big[{\cal R}_l+\sum_{j\neq l}(\Omega_{jl}+\text{i}\gamma_{jl}){\rho}^{(j)}_{ge}\big],
	\end{split}\label{coherence}\\
	\begin{split}
	\dot{\rho}^{(l)}_{ee}=&-2\gamma {\rho}^{(l)}_{ee}+2\text{Im}[{\cal R}_l^*{\rho}^{(l)}_{ge}]\\
	&+2\text{Im}\big[\sum_{j\neq l}(\Omega_{jl}-\text{i}\gamma_{jl}){\rho}^{(l)}_{ge}({\rho}^{(j)}_{ge})^*\big],
	\end{split}\label{excitations}
	\end{align}
\end{subequations}
where we define $\rho_{ge}^{(l)}=\langle \hat{\sigma}{}_l^{-}\rangle$ and $\rho_{ee}^{(l)}=\langle \hat{\sigma}{}_l^{ee}\rangle$.
The summation terms in Eqs.~\eqref{SpinEqs} describe light-mediated interactions between a discrete atoms at fixed points $l$ and $j$ in a lattice. The DD interaction terms $\Omega_{jl}$ and $\gamma_{jl}$ depend on the relative positions between the atoms in the lattice, and are given by the real and imaginary part of the dipole radiation kernel, Eq.~\eqref{Gdef}, 
\begin{equation} \label{dipolekernel}
\begin{split}
&\Omega_{jl}= \frac{1}{\hbar \epsilon_0}\text{Re}\left[\textbf{d}_{ge}^*\cdot \mathsf{G}(\textbf{r}_j-\textbf{r}_l)\textbf{d}_{ge}\right],\\
&\gamma_{jl}=\frac{1}{\hbar \epsilon_0}\text{Im}\left[\textbf{d}_{ge}^*\cdot \mathsf{G}(\textbf{r}_j-\textbf{r}_l)\textbf{d}_{ge}\right],
\end{split}
\end{equation}
where $\gamma_{jj}=\gamma =\mathcal{D}^2 k^3/(6\pi\epsilon_0\hbar)$ is the single-atom linewidth.
DD interactions result in recurrent and correlated light scattering between the atoms with a strong collective response from the array. The contact term of the scattered light field in Eq.~\eqref{Gdef} is inconsequential in the atomic interaction coefficients of Eqs.~\eqref{SpinEqs}
and is the origin of the local field shift of light inside the medium~\cite{Ruostekoski1997b}. 
In the absence of DD interactions, Eqs.~\eqref{SpinEqs} reduce to the usual independent-atom optical Bloch equations. 
Mean-field equations based on related principles as those in Eqs.~\eqref{SpinEqs} have also been used to describe systems with and without spatial fluctuations~\cite{Lee16,Sutherland_satur,Lee17,Machluf2018,Kramer2015a}.
Other transitions can be included, such as the $m=\pm1$ states of the $\ket{J=0,m=0} \rightarrow \ket{J=1,m=0,\pm1}$ transition, and the corresponding general form of Eqs.~\eqref{SpinEqs} with arbitrary internal atomic levels is given in Ref.~\cite{Lee16}, with the specific case of three-level system simulated in Ref.~\cite{Machluf2018}.
However, this quickly increases the complexity of the system. By application of a magnetic field, the $m=\pm1 $ states can be tuned far off-resonance so an effective two-level system can be obtained with the $m=0$ states.

\subsubsection{Low light intensity}

In the limit of LLI, atoms occupy the ground state, with changes to the coherence, $\rho_{ge}$, linearly proportional to the incident light field amplitude, $ \boldsymbol{\mathcal{E}}^+$. 
The LLI limit~\cite{Ruostekoski1997a, Lee16} then constitutes deriving the equations first order in light field amplitude by keeping the terms that include at most one of the amplitudes $\rho_{ge}$ or $ \boldsymbol{\mathcal{E}}^+$, and no $\rho_{ee}$.
Equations~\eqref{SpinEqs} then reduce to a linear set of equations describing $N$ dipole-coupled oscillators
\begin{equation}
\begin{split}
&\dot{\rho}_{ge}^{(l)} = \text{i}\sum_j \mathcal{H}_{jl}\rho_{ge}^{(j)} +\text{i}\mathcal{R}_l,\\
&\mathcal{H}_{jl} = (\Delta+\text{i}\gamma)\delta_{jl}+(\Omega_{jl}+\rm{i}\gamma_{jl})(1-\delta_{jl}).
\end{split}
\end{equation}
The LLI collective excitation eigenmodes are given by the eigenmodes of $\mathcal{H}$, which satisfy biorthogonality relations, with complex eigenvalues $\delta_{\alpha}+\text{i}\upsilon_{\alpha}$, where $\delta_{\alpha}$ and $\upsilon_{\alpha}$ describe the collective line shift (from the resonance of the isolated atom) and linewidth, respectively~\cite{Rusek96,Sokolov2011,JenkinsLongPRB,Jenkins2012a,Castin13,Jenkins_long16,Facchinetti16,Lee16}.
Modes with $\upsilon_{\alpha}>\gamma$ ($\upsilon_{\alpha}<\gamma$) are termed superradiant (subradiant). 
For an infinite lattice, the eigenmodes of the system are described by plane waves with wavevector $\textbf{q}$. The eigenvalues are now $\tilde{\Omega}(\textbf{q})+\rm{i}[\tilde{\gamma}(\textbf{q})+\gamma]$ where 
\begin{equation}\label{FourierTransforms}
\begin{split}
&\tilde\Omega(\textbf{q})=\sum_{j\neq l}\Omega_{jl}e^{\text{i}\textbf{q}\cdot\textbf{r}_{j}},\quad\tilde{\gamma}(\textbf{q})=\sum_{j\neq l}\gamma_{jl}e^{\text{i}\textbf{q}\cdot\textbf{r}_{j}},
\end{split}
\end{equation}
are the Fourier transforms of the real and imaginary parts of the dipole kernel, Eq.~\eqref{dipolekernel}, respectively (excluding the self-interaction $j=l$). 
The plane waves are given by
\begin{align}
v^{(+)}_{{\bf q}}({\bf r}_l) &=A_{{\bf q}}\cos(\textbf{q}\cdot\textbf{r}_l), \\ 
v^{(-)}_{{\bf q}}({\bf r}_l) & =A_{{\bf q}}\sin(\textbf{q}\cdot\textbf{r}_l),
\end{align}
where $A_{{\bf q}} = \sqrt{2/N}$ except for $A_{{\bf q} = 0, (\pi/a, \pi/a)} = 1/\sqrt{N}$. For the infinite system, the two-level transition resonance wavelength defines the light cone, $k a/\sqrt{2}$, where any mode with $|\textbf{q}|> k/\sqrt{2}$ results in $\tilde{\gamma}(\textbf{q}) = -\gamma$ and the corresponding mode is then completely dark.

\subsubsection{Solutions to the mean-field equations}

We now solve the dynamics of Eqs.~\eqref{SpinEqs} to obtain the steady-state solutions by considering uniform level shifts, $\Delta_l = \Delta$. For the incident plane with the wave vector ${\bf k}$, 
a phase varying Rabi frequency, ${\cal R}_l={\cal R} e^{i\textbf{q}\cdot\textbf{r}_l}$, can be obtained  by tilting the angle of incidence, such that $\textbf{q}=\textbf{k}-(\hat{\textbf{z}}\cdot\textbf{k})\hat{\textbf{z}}$. 
A general solution to Eqs.~\eqref{SpinEqs} is then given by $\rho^{(l)}_{ee}=\rho_{ee}$ and $\rho_{ge}^{(l)}=\rho_{ge}e^{i\textbf{q}\cdot\textbf{r}_l}$, with 
\begin{equation}\label{SScoherence}
\begin{split}
\rho_{ge}=\frac{\text{i}{\cal R} Z}{\text{i}\left[\Delta-Z\tilde\Omega(\textbf{q})\right]-\left[\gamma-Z\tilde{\gamma}(\textbf{q})\right]},
\end{split}
\end{equation} 
where
\begin{equation}
Z=2\rho_{ee}-1.
\end{equation}
However, modes with $\textbf{q}$ lying near to or outside the light cone cannot directly be excited by incident light due to the rapid phase variation required, and instead must be driven by applying symmetry-breaking fields to the lattice~\cite{parmee2020}. Such a symmetry-breaking level shifts could be generated, for example, by ac Stark shifts~\cite{gerbier_pra_2006} of lasers. 

The population difference, $Z$, obeys the cubic equation
\begin{equation}\label{SScubic}
\begin{split}
p(Z)&=\big[\tilde{\gamma}(\textbf{q})^2+\tilde\Omega(\textbf{q})^2\big]Z^3+(\Delta^2+\gamma^2)\\
&+\big[\tilde{\gamma}(\textbf{q})^2+\tilde\Omega(\textbf{q})^2-2\Delta \tilde\Omega(\textbf{q})-2\gamma \tilde{\gamma}(\textbf{q})\big]Z^2\\
&+\big[\Delta^2+\gamma^2+2|{\cal R}|^2-2\Delta \tilde\Omega(\textbf{q}) -2\gamma \tilde{\gamma}(\textbf{q})\big]Z\\
&=0.
\end{split}
\end{equation}
When $|\mathcal{R}|^2\approx0$, Eq.~\eqref{SScubic} admits the solution $\rho_{ee}=0$, with the coherence, Eq.~\eqref{SScoherence}, now describing the LLI eigenmode of the infinite system with a wave vector $\textbf{q}$. 
However, for nonzero incident fields, Eq.~\eqref{SScubic} can have up to three real solutions, of which two are dynamically stable, resulting in optical bistability.
Cases where solutions become unstable can result in the emergence of a typically rich phase diagram of different solutions, exhibiting a dependence on the intensity and laser frequency~\cite{parmee2020}.

%%%%%%%%%%%%%%%%%%%%%%%%%%%%%%%%%%%%%%%%%%%%%%%%%%%%%%%%%%%%%
\section{Bistability in arrays of atoms}\label{2Darrays}

\subsection{General formalism}

We now establish the formalism used to determine the parameter ranges where bistability is possible for an array of atoms. To do this, we substitute $\rho^{(l)}_{ee}=\rho_{ee}$ and $\rho_{ge}^{(l)}=\rho_{ge}e^{i\textbf{q}\cdot\textbf{r}_l}$ into Eqs.~\eqref{SpinEqs}, and rewrite as
\begin{subequations}\label{SpinEqsMacro3}
	\begin{align}
	\dot{\rho}_{ge}=&\left(\text{i}\Delta-\gamma\right) \rho_{ge} - \text{i}(2\rho_{ee}-1) \mathcal{R}_{\rm eff}, \label{SEq}\\
	\dot{\rho}_{ee}=&-2\gamma \rho_{ee}+2\text{Im}\left[\mathcal{R}_{\rm eff}^*\rho_{ge}\right],\label{Zeq}
	\end{align}
\end{subequations}
where we have defined
\begin{equation}\label{Eq:EDef}
{\cal R}_{\rm eff}= {\cal R}+[\tilde{\Omega}(\textbf{q})+\text{i}\tilde {\gamma}(\textbf{q})]\rho_{ge},
\end{equation}
which is the total external electric field (incident plus scattered field from all the other atoms, given in terms of the Rabi frequency) driving an arbitrary atom $l$ in the ensemble. 
Solving Eqs.~\eqref{SpinEqsMacro3} gives the coherence and excited level population in terms of $\mathcal{R}_{\rm{eff}}$~\cite{parmee2020},
\begin{subequations}\label{Eq:EOMSolns}
	\begin{align}
	\rho_{ge} &={\cal R}_{\rm eff}\frac{-\Delta+\text{i}\gamma}{\Delta^2+\gamma^2+2|{\cal R}_{\rm eff}|^2},\label{Eq:Coherence}\\
	\rho_{ee} &=\frac{|{\cal R}_{\rm eff}|^2}{\Delta^2+\gamma^2+2|{\cal R}_{\rm eff}|^2} \label{Eq:Excitations}.
	\end{align}
\end{subequations}
These solutions have a similar form to the solutions of the optical Bloch equations,
but now with the Rabi frequency, $\mathcal{R}$, replaced by $\mathcal{R}_{\rm{eff}}$.
Using Eq.~\eqref{Eq:Coherence} to eliminate $\rho_{ge}$ from Eq.~\eqref{Eq:EDef} gives
\begin{equation}\label{Eq:yequation}
\begin{split}
\mathcal{R}&={\cal R}_{\rm eff}+{\cal R}_{\rm eff}\frac{2C(\Delta^2+\gamma^2)}{\Delta^2+\gamma^2+2|\mathcal{R}_{\rm eff}|^2},
\end{split}
\end{equation}
where we have defined the \emph{cooperativity parameter}~\cite{parmee2020},
\begin{equation}\label{Eq:GsumDefinition}
\begin{split}
C &= \frac{1}{2}\frac{\tilde{\Omega}(\textbf{q})+\text{i}\tilde{\gamma}(\textbf{q})}{\Delta+\text{i}\gamma},
\end{split}
\end{equation} 
which is a measure of the collective behavior in the array and plays an important role in describing bistability.
Finally, by taking the absolute value of both sides of Eq.~\eqref{Eq:yequation}, we obtain
\begin{equation}\label{Eq:Modyequation}
\begin{split}
\frac{I}{I_{\rm sat}}&=\frac{2|\mathcal{R}_{\rm eff}|^2}{\gamma^2}\frac{1}{(\eta^2+2|\mathcal{R}_{\rm eff}|^2)^2}\biggr\{4\eta^4\text{Im}[C]^2\\
&\phantom{=======}+\left[\eta^2(1+2\text{Re}[C])+2|\mathcal{R}_{\rm eff}|^2\right]^2\biggr\},
\end{split}
\end{equation}
where
\begin{equation}\label{Eq:alphabeta}
\begin{split}
\eta^2=\Delta^2+\gamma^2,
\end{split}
\end{equation}
and $I/I_{\rm{sat}}$ is given by Eq.~\eqref{intensity} (where the intensity is now the same for all sites, $l$).
Equation~\eqref{Eq:Modyequation} is a cubic equation in $|\mathcal{R}_{\rm eff}|^2$, with either one or two dynamically stable real solutions, and the bistability region found when the discriminant is zero as a function of $I/I_{\rm{sat}}$ and $\Delta$. 
Equation~\eqref{SScubic} can also be used to determine bistability in the system. 
However, introduction of the effective field and cooperativity parameter in Eq.~\eqref{Eq:Modyequation} recasts the equations in the same notation used for bistability in cavity systems~\cite{Bonifacio1978}, making the two systems easier to compare.

For large enough lattice spacings, there is only a single solution to Eq.~\eqref{Eq:Modyequation}, and hence no bistability for any intensity or detuning. To determine the minimal lattice spacing for the array to support bistability, we consider $I/I_{\rm sat}$ as a function of $|\mathcal{R}_{\rm eff}|^2$ in Eq.~\eqref{Eq:Modyequation}, and find the lattice spacing where two minima develop, which involves solving $dI/ d |{\cal R}_{\rm eff}|^2 = 0$, explicitly given by
\begin{equation}\label{Eq:IntensityCubic}
\begin{split}
&4|{\cal R}_{\rm eff}|^2\left(\eta^2+2|{\cal R}_{\rm eff}|^2\right)\left(\eta^2+2|{\cal R}_{\rm eff}|^2+2\eta^2\text{Re}[C]\right)\\
&+\left(\eta^2-2|{\cal R}_{\rm eff}|^2\right)\left[\left(\eta^2+2|{\cal R}_{\rm eff}|^2+2\eta^2\text{Re}[C]\right)^2\right]\\
&+4\eta^4\text{Im}[C]^2\left(\eta^2-2|{\cal R}_{\rm eff}|^2\right)=0.
\end{split}
\end{equation}

\subsection{Analytic bistable solutions}

For closely-packed arrays where $\tilde{\Omega}(\textbf{q}),\tilde{\gamma}(\textbf{q})  \gg \Delta,\gamma$, Eq.~\eqref{Eq:Modyequation} has two well-separated minima and the bistable solutions can be approximated.
For low intensities [$\tilde{\Omega}(\textbf{q})/\gamma,\tilde{\gamma}(\textbf{q})/\gamma  \gg I/I_{\rm{sat}}$], we obtain $|\mathcal{R}_{\rm{eff}}|$ from Eq.~\eqref{Eq:Modyequation} using 
\begin{equation}
\begin{split}
&(\eta^2+2|\mathcal{R}_{\rm eff}|^2+2\eta^2\text{Re}[C])^2\approx4\eta^4\text{Im}[C]^2\\
&+4|\mathcal{R}_{\rm{eff}}|^2\eta^2(2\text{Re}[C]+1)+(1+2\text{Re}[C])^2\eta^4,
\end{split}
\end{equation}
and use Eq.~\eqref{Eq:yequation} to obtain the phase.
For high intensities [$I/I_{\rm{sat}}  \gg \tilde{\Omega}(\textbf{q})/\gamma,\tilde{\gamma}(\textbf{q})/\gamma,\Delta^2/\gamma^2$], $\mathcal{R}_{\rm eff}$ is found from Eq.~\eqref{Eq:yequation} by ignoring the $\Delta^2+\gamma^2$ term in the denominator.
The approximate solutions for low and high intensities are then, respectively, given by
\begin{subequations}
	\label{CoopandSingleAtomSoln}
	\begin{align}
	\begin{split}
	{\cal R}_{\rm eff}^{\rm coop} &=\frac{\sqrt{2}\mathcal{R}}{2C+1} \biggr[1-\frac{4|\mathcal{R}|^2}{\eta^2|2C+1|^2}\\
	&\phantom{==}+\sqrt{1+(1-|2C|^2)\frac{8|\mathcal{R}|^2}{\eta^2|2C+1|^4}}\biggr]^{-1/2},
	\end{split} \label{CoopSoln}\\
	\begin{split}
	{\cal R}_{\rm eff}^{\rm SA} &= \frac{\mathcal{R}}{2}\biggr[1-\frac{2\text{i}\eta^2\text{Im}[C]}{|\mathcal{R}|^2}\\
	&\phantom{==}+\sqrt{1-\frac{4\eta^2\text{Re}[C]}{|\mathcal{R}|^2}-\left(\frac{2\eta^2\text{Im}[C]}{|\mathcal{R}|^2}\right)^2}\biggr] \label{SingleAtomSoln},
	\end{split}
	\end{align}
\end{subequations} 
where we have labeled the solutions as the ``cooperative'' and ``single-atom'' due to their very different responses to the incident light\footnote{Note the cooperative solution presented here is more accurate approximation than the solution presented in our previous work~\cite{parmee2020}.}, in an analogy with a similar terminology in optical cavities~\cite{Bonifacio1978}.
For the cooperative solution, Eq.~\eqref{CoopSoln}, the atoms behave collectively, creating a field that counteracts the incident light and resulting in the atoms absorbing strongly, especially at higher atom densities.
This is demonstrated most clearly in the LLI limit, where  $\mathcal{R}^{\rm{coop}}_{\rm{eff}}\approx\mathcal{R}/(2C+1)$, where we can see how the effective field scales inversely with $C$, with strongly collective behavior resulting in a small $\mathcal{R}_{\rm{eff}}$.
For the single-atom solution, Eq.~\eqref{SingleAtomSoln}, the atoms now saturate and absorption is weak, with the medium becoming transparent as the atoms react to the incident light almost independently. 
The effective field scales linearly with the incident field for high intensities, where $\mathcal{R}^{\rm{SA}}_{\rm{eff}}\approx\mathcal{R}$ and there is no dependence on $C$ as collective behavior between the atoms is lost. 

The cooperative and single-atom solutions only describe the system response for the intensity ranges
\begin{subequations}
	\label{IntensityBounds}
	\begin{align}
	&\frac{I}{I_{\rm{sat}}} <\frac{\eta^2}{4\gamma^2}\frac{|2C+1|^4}{|2C|^2-1} \label{CooperativeIntensityBound},\\
	& \frac{4\eta^2}{\gamma^2}\left(\text{Re}[C] + |C|\right) \label{SingleAtomIntensityBound} < \frac{I}{I_{\rm{sat}}},
	\end{align}
\end{subequations} 
respectively,
and serve as approximate lower and upper intensity bound of the bistability region. However, a more accurate analytic approximation for the upper intensity bound can be found using Eq.~\eqref{Eq:IntensityCubic} in the limit that $\tilde{\Omega}(\textbf{q}),\tilde{\gamma}(\textbf{q})  \gg \Delta, \gamma$ by expanding the cubic solutions to Eq.~\eqref{Eq:IntensityCubic} about small $\eta^2$ that yields
\begin{equation}\label{Eq:IntensityRange}
\begin{split}
&\frac{4\eta^2}{\gamma^2}\left(\text{Re}[C] +|C|\right) <\frac{I}{I_{\rm sat}}<\frac{\eta^2}{\gamma^2}|C+1|^2.
\end{split}
\end{equation}

\subsection{Analytic thresholds}

Finding the thresholds for bistability by solving Eq.~\eqref{Eq:IntensityCubic} can usually only be done numerically. However, there are two cases where analytic solutions can be easily obtained~\cite{parmee2020}.
The first case is for real $C$ 
when $\Delta/\gamma = \tilde{\Omega}(\textbf{q})/\tilde{\gamma}(\textbf{q})$, where Eq.~\eqref{Eq:IntensityCubic} gives a threshold of
\begin{equation}\label{thresh}
\tilde{\gamma}(\textbf{q})>8\gamma.
\end{equation}
The second case is for imaginary $C$ 
when $\Delta/\gamma = -\tilde{\gamma}(\textbf{q})/\tilde{\Omega}(\textbf{q})$, where
Eq.~\eqref{Eq:IntensityCubic} forms a cubic equation 
where two positive real solutions and bistability are possible when
\begin{equation}\label{Disc}
\begin{split}
[\tilde{\Omega}(\textbf{q})]^2>27\gamma^2.
\end{split}
\end{equation}

\subsubsection{Two atom bistability}

The analytic results reveal high density thresholds for the emergence of optical bistability. This can already be seen in the simplest possible case of two closely spaced atoms within the mean-field approximation~\cite{Bettles2020} and under uniform illumination.
The condition $\tilde{\gamma}=\gamma_{12}>8\gamma$ cannot be met as $\gamma_{12}\rightarrow \gamma$ in the limit of zero separation. However, the threshold $\tilde{\Omega}=\Omega_{12} > \sqrt{27} \gamma$ can be satisfied, and a simple dimensional analysis for $\Omega_{12}\sim 1/(ka)^3$ yields the threshold $ka\alt1$. This value also equals the separation required for the collective shift
to exceed the single-atom linewidth $\Omega_{12}\agt \gamma$;  a condition at which correlations due to light-mediated interactions lead to substantial deviations from standard continuous medium optics~\cite{Javanainen2014a}.
A more accurate calculation gives the bistability threshold $ka \alt0.94 $ (corresponding to a lattice spacing of $a \alt0.15\lambda $) and $ka \alt0.63 $ ($a \alt0.10\lambda $) for atoms polarized parallel or perpendicular to the separation axis, respectively. 

\subsubsection{Arrays of atoms}

Analytic expressions for the optical bistability can be obtained for planar arrays  for $\Delta/\gamma =  \tilde\Omega(\textbf{q}) /\tilde{\gamma}(\textbf{q}) $, when the solutions no longer depend on $\tilde\Omega(\textbf{q})$.
The collective radiative linewidth for the uniform LLI eigenmode with atomic dipoles polarized in the lattice plane
 $\tilde{\gamma}(\textbf{0})+\gamma$ has a simple analytic form~\cite{CAIT} (see also Ref.~\cite{Facchinetti18}) 
 \begin{equation}\label{Analytic}
\begin{split}
\tilde{\gamma}(\textbf{0})=-\gamma+\frac{3\pi\gamma}{(ka)^2}.
\end{split}
\end{equation}
The bistability threshold $\tilde{\gamma}(\textbf{0}) >8\gamma$ is then met when $k a < (\pi/3)^{1/2}$ ($a \alt 0.163\lambda$). Note that for the uniform LLI for which the
dipoles point normal to the array plane, $\tilde{\gamma}(\textbf{0})  = -\gamma$~\cite{Facchinetti16}, and so $\tilde{\gamma}(\textbf{0}) \not> 8\gamma$ and bistability is not possible.
Similar bistability threshold can also be evaluated analytically for an infinite 1D chain of atoms~\cite{parmee2020}, giving $k a < \pi/6$ or $k a < \pi/12$ for 
dipoles parallel and perpendicular to the chain, respectively.

\subsubsection{Analogy with optical cavities}

Analytic solutions for $\Delta/\gamma = \tilde{\Omega}(\textbf{q})/\tilde{\gamma}(\textbf{q})$, for which case the cooperativity parameter becomes real, with $C=\tilde{\gamma}(\textbf{q})/2\gamma$, provide a special case where the precise analogy of the optical bistability in atomic arrays and optical cavities~\cite{Lugiato1984,bonifacio1976,Bonifacio1978,Carmichael1977,Agrawal79,Carmichael1986a} can be established (see Appendix~A). 
The cooperativity parameter $C=Ng^2/2\gamma\kappa$ for an optical cavity~\cite{Bonifacio1978,CarmichaelVol2}, as well as 
the incident and total fields then satisfy exactly the same formulaic relation, with the condition $2C>1$ corresponding to the strong coupling regime of optical cavities. Moreover, in the limit of a vanishing $\kappa$,  $C=N g/2\gamma$, and $\tilde{\gamma}(\textbf{q})$ takes the role of the atom-cavity coupling coefficient $Ng$.
Large $C$ in optical cavities represents many recurrent scattering events of an atom with light reflecting between the cavity mirrors, while in arrays in free space
it represents recurrent scattering events with neighboring atoms at high densities with $ka \sim 1$.

\subsection{Numerical solutions}

\subsubsection{Uniform incident field}\label{unibistability}
\begin{figure}
	\vspace{-0.4 cm}
	\includegraphics[width=\widthscale\columnwidth]{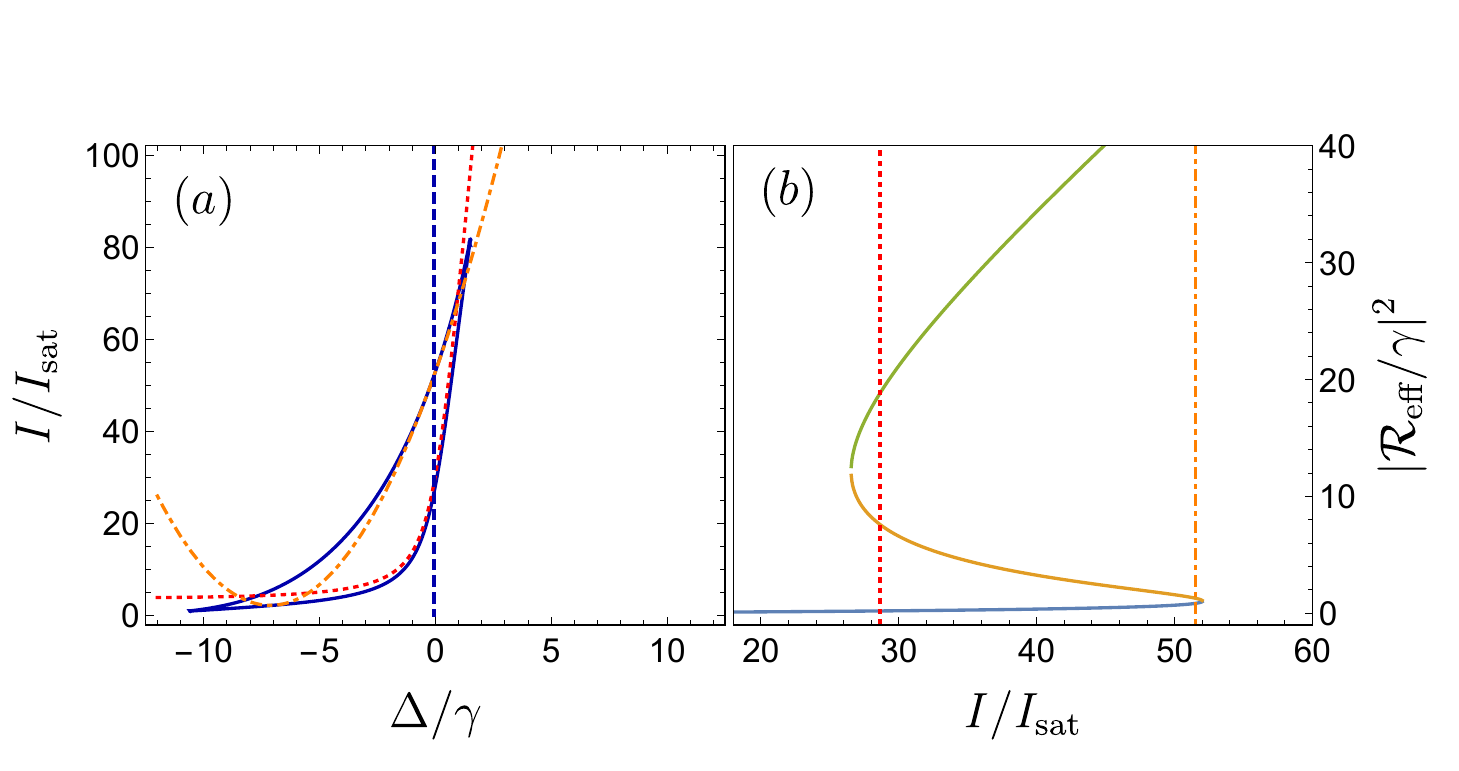}
	\vspace{-0.7 cm}
	\caption{
	Bistability of two atoms separated by $0.1\lambda$ with the dipoles polarized parallel to the separation axis. (a) The region of bistability from Eq.~\eqref{Eq:Modyequation} (solid blue line) and (b) $|{\cal R}_{\rm eff}/\gamma|^2$ for $\Delta/\gamma = -\gamma_{12}/\Omega_{12}$ [blue dashed line in (a)]. The intensity thresholds, Eq.~\eqref{Eq:IntensityRange} (red-dotted and orange-dot-dashed line) are also shown in both (a,b).
	}
	\label{Bistability2Atoms}
\end{figure}

\begin{figure}
	\hspace*{0cm}
	\includegraphics[width=\widthscale\columnwidth]{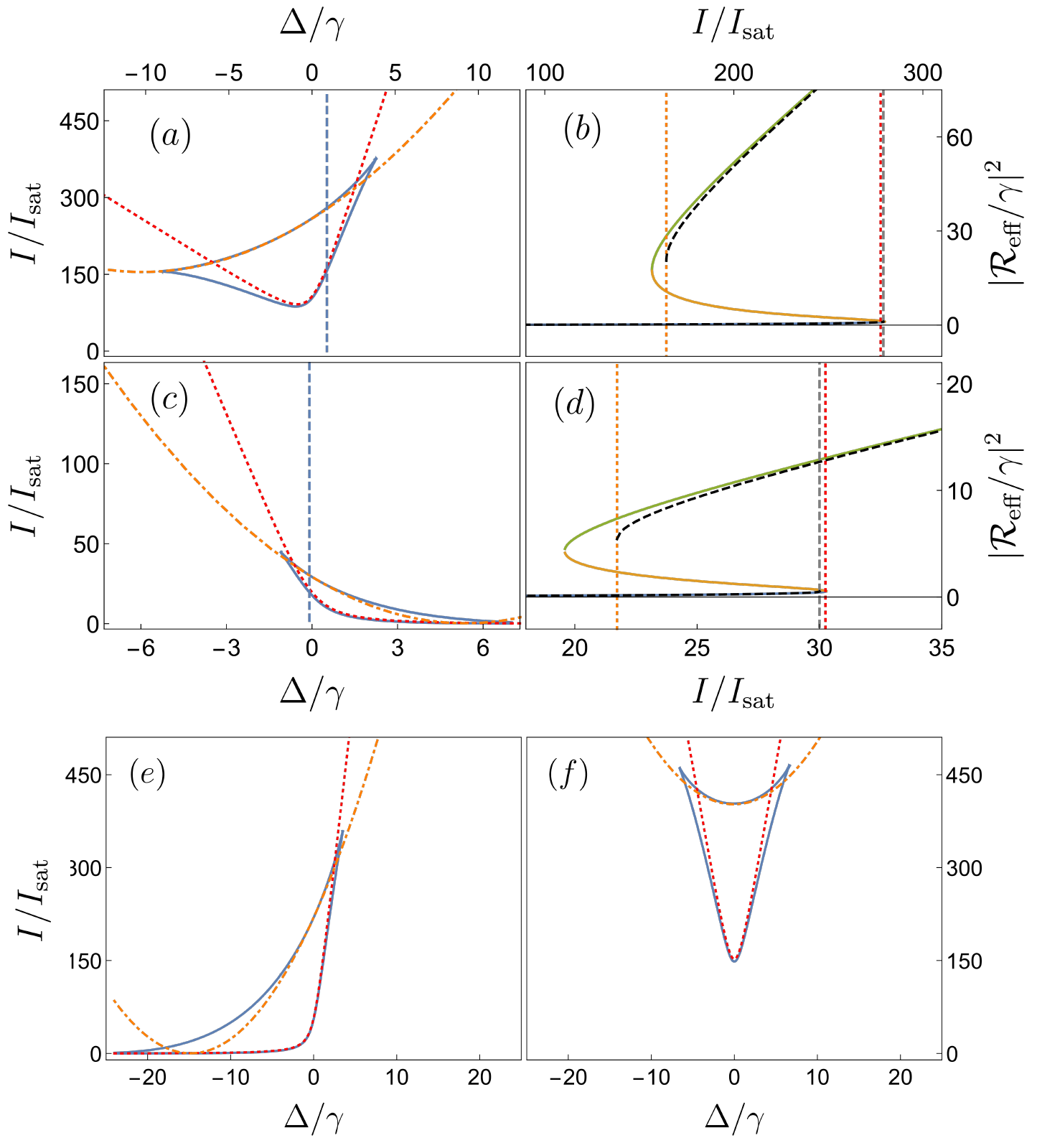}
	\vspace{-0.7cm}
	\caption{Optical bistability in a planar array of atoms. 
		(a) Region of bistability from Eq.~\eqref{Eq:Modyequation} (solid blue line) for $\textbf{q}=\textbf{0}$, $a=0.1\lambda$ and $\hat{\textbf{e}}=(1,1,0)/\sqrt{2}$. (b) $|{\cal R}_{\rm eff}/\gamma|^2$ for $\Delta/\gamma = \tilde\Omega(\textbf{0})/\tilde\gamma(\textbf{0}) $ [blue dashed line in (a)] with approximate cooperative (lower black dashed line) and single-atom (upper black dashed line) solutions, Eqs.~\eqref{CoopandSingleAtomSoln}. 
		(c) Bistability region and (d) $|{\cal R}_{\rm eff}/\gamma|^2$ for $\Delta/\gamma = -\tilde\gamma(\textbf{q})/\tilde\Omega(\textbf{q})$ [blue dashed line in (c)] with approximate cooperative and single-atom solutions for the mode $\textbf{q} = (\pi/a,\pi/a)$.
		(e) Bistability region crossing the light cone for the mode $\textbf{q} = (\pi/4a,\pi/4a)$ at the lattice spacings $a=0.17\lambda$ and (f) $a=0.18\lambda$.
		In all plots, the intensity thresholds Eqs.~\eqref{Eq:IntensityRange} (red-dotted and orange-dot-dashed lines) are shown. In (b,d), the cooperative solution intensity threshold  Eq.~\eqref{CooperativeIntensityBound} (gray dashed line) is also shown.
	}
	\vspace{0cm}
	\label{Fig:S-curve}
\end{figure}

For a uniformly illuminated two-atom system, the bistability region, obtained from Eq.~\eqref{Eq:Modyequation}, is shown 
in Fig.~\ref{Bistability2Atoms}(a), and the emergence of multiple solutions of ${\cal R}_{\rm eff}/\gamma$ as a function of $I/I_{\rm sat}$ in Fig.~\ref{Bistability2Atoms}(b).
Similar plots for a planar array are shown in
Figs.~\ref{Fig:S-curve}(a,b), with the corresponding cooperative and single-atom approximations [Eqs.~\eqref{CoopSoln} and~\eqref{SingleAtomSoln} respectively].
The effective field $\mathcal{R}_{\rm{eff}}$ is small for the cooperative solution due to the collective nature of the solution, while the single-atom solution scales linearly with the incident field at high intensities as the atoms saturate and behave independently. 
Upon varying the detuning or intensity in experiments, the system will jump between the cooperative and single-atom solutions within the region of bistability. This will result in hysteresis in the transmission properties of the lattice, addressed in more detail in Sec.~\ref{transmission}.
As the lattice spacing is increased, the size of the bistability region shrinks as recurrent scattering between the atoms, and hence $C$, decreases, with bistability completely lost for a lattice spacing of $a\simeq0.165\lambda$,
very close to the analytic value $k a = (\pi/3)^{1/2}$, or $a\simeq0.163\lambda$.

\subsubsection{Nonuniform incident fields}

\begin{figure}
	\hspace*{0cm}
	\includegraphics[width=\widthscale\columnwidth]{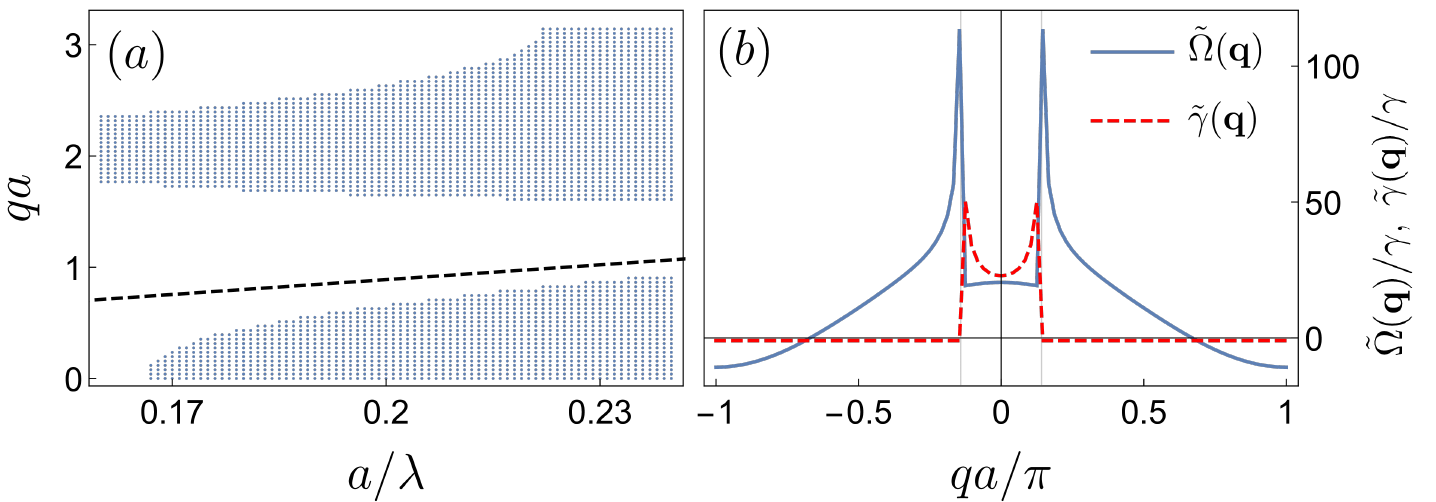}
	\vspace{-0.7cm}
	\caption{Loss of optical bistability with lattice spacing. 
	(a) Modes that lose bistability (blue dots) as a function of lattice spacing for $\textbf{q}=q(1,1)$, which are divided into two groups by the light cone (black dashed line). 
	(b) Collective radiative couplings $\tilde{\Omega}(\textbf{q})$ and $\tilde{\gamma}(\textbf{q})$ [Eqs.~\eqref{FourierTransforms}] for a lattice with $ka = 0.2\pi$. For $\textbf{q}=\textbf{0}$, Eq.~\eqref{Analytic} gives $\tilde{\gamma}(\textbf{0}) \simeq 22.9\gamma$.
	}
	\vspace{0cm}
	\label{BSthreshold2D}
\end{figure}

Numerical solutions to $\tilde{\Omega}(\textbf{q})$ and $\tilde{\gamma}(\textbf{q})$ from Eqs.~\eqref{FourierTransforms} can be obtained efficiently by computing the sum in momentum space~\cite{Perczel2017,Abajo07,Asenjo_prx,Antezza2009}.
Figures~\ref{Fig:S-curve}(c,d) show the bistability region for a checkerboard-pattern excitation $\textbf{q} = (\pi/a,\pi/a)$ for which the corresponding LLI eigenmode is subradiant.
Bistability occurs within a smaller detuning range and at lower intensities than for the superradiant $\textbf{q}=\textbf{0}$ mode. We find this to be a general feature for all subradiant modes, due in part to a smaller $\tilde{\gamma}(\textbf{q})$ and hence smaller $C$. The emergence of bistability is also related to the nonlinear response of the LLI eigenmodes, which has been shown to depend on their linewidth~\cite{Williamson2020}, with a smaller linewidth resulting in lower intensities for a nonlinear response.
As for $\textbf{q}=\textbf{0}$, the bistability region decreases in size with increased lattice spacing. However, some modes cross the light cone which leads to large changes in the bistability region. An example of this is shown in Fig.~\ref{BSthreshold2D}(e-f) for the mode $\textbf{q} = (\pi/4a,\pi/4a)$. At small lattice spacings, this mode lies outside the light cone, with a highly asymmetric bistability region at low intensities. However, the mode crosses inside the light cone at the lattice spacing $a\simeq0.177\lambda$, with $\tilde{\gamma}(\textbf{q})$ increasing drastically, resulting in a highly symmetric bistability region at much higher intensities. 

The lattice spacing where optical bistability vanishes depends on the corresponding LLI eigenmode, shown in Fig.~\ref{BSthreshold2D}(a). 
The loss of bistability for modes inside the light cone is well described by the threshold $\tilde{\gamma}(\textbf{q})> 8\gamma$, with $\tilde{\gamma}(\textbf{q})$ plotted in Fig.~\ref{BSthreshold2D}(b). At the light cone, $\tilde{\gamma}(\textbf{q})$ becomes large, so there is always a mode that satisfies $\tilde{\gamma}(\textbf{q}) > 8\gamma$, and the light cone therefore acts as an upper bound for bistability loss in Fig~\ref{BSthreshold2D}(a).
Outside the light cone, $\tilde{\gamma}(\textbf{q})=-\gamma$, and loss of bistability is determined purely by changes in $\tilde{\Omega}(\textbf{q})$.
For other orientations of the atomic transition within the lattice plane, e.g., $\hat{\textbf{e}}=(1,0,0)$, the variation of the collective linewidth $\tilde{\gamma}(\textbf{q})$ with $\textbf{q}$ changes, and $\tilde{\gamma}(\textbf{q})$ no longer diverges at the light cone. This means $\tilde{\gamma}(\textbf{q}) > 8\gamma$ is not always satisfied for modes at the light cone and bistability of these modes do not always persist to larger lattice spacings or undergo sharp changes if the mode crosses the light cone.
However, in general, bistability of subradiant modes persists to larger lattice spacings than superradiant modes.
%

%%%%%%%%%%%%%%%%%%%%%%%%%%%%%%%%%%%%%%%%%%
\section{Transmission}\label{transmission}

The bistable solutions, Eqs.~\eqref{CoopandSingleAtomSoln}, exhibit very different optical responses. We now analyze how this modifies the transmission of light through the array and focus on uniform incident fields normal to the array with $\textbf{q}=\textbf{0}$.

Experiments on transmission through a $14\times14$ optical lattice of ${}^{87}\text{Rb}$ atoms with near unit filling have recently been performed in the LLI limit~\cite{Rui2020}. 
The lattice spacing of $a=0.68\lambda$ exhibits a subradiant eigenmode with a spatially uniform phase profile and (in the absence of position uncertainty) collective linewidth $\gamma+\tilde{\gamma} \simeq 0.52\gamma$ [Eq.~\eqref{Analytic}] that was driven by the incident field and observed in a narrowed transmission resonance for light 
in a dramatic demonstration of subradiance. While generating subradiance does not necessarily require very dense atomic
ensembles~\cite{Guerin_subr16}, the advantage of the lattice system is that a significant fraction of the atoms can occupy the same subradiant eigenmode, providing at the same time also a close analogy to subradiance in metasurfaces of fabricated resonators~\cite{Jenkins17}. In addition to optical lattice experiments, atoms in optical tweezer arrays are now illuminated with resonant light~\cite{Glicenstein2020}.

In our bistability study, we analyze the light transmission beyond the LLI, while coupling to the same LLI eigenmode as in
the transmission experiment of Ref.~\cite{Rui2020}. The smaller lattice spacing $a=0.1\lambda$, however, transforms this LLI eigenmode to
a superradiant one with the collective linewidth $\gamma+\tilde{\gamma}\simeq23.9\gamma$ [Eq.~\eqref{Analytic}],
but surprisingly we find that many of the similar transmission properties can persist well beyond the limit of LLI.
Different spacings can be experimentally achieved by different laser angles and, e.g., by different atomic species and transitions, such as the $\mbox{}^3P_0\rightarrow \mbox{}^3D_1$ transition in $^{88}$Sr~\cite{Olmos13}, exhibiting a resonance wavelength of $\lambda\simeq 2.6\mu$m and achievable spacing of 206.4nm, with the effective lattice spacing $a\simeq 0.08\lambda$. 
We also analyze hysteresis of light transmission by varying the detuning over a timescale of $\tau = 250/\gamma$ during the dynamics. For instance, for Rb atoms, this gives a timescale of $\tau = 1.3\times10^{-5}$s -- significantly shorter than typical trapping times of atoms in the optical lattices.

\subsection{Extinction and reflectivity}

We consider coherent light transmission through the array with the transmission and reflection amplitudes,
\begin{equation}\label{Eq:TransmissionAmplitude}
t=\frac{\int\hat{\textbf{e}}\cdot\textbf{E}^+(\textbf{r})dS}{\int\hat{\textbf{e}}\cdot\boldsymbol{\mathcal{E}}{}^+(\textbf{r})dS},\quad r=\frac{\int\hat{\textbf{e}}\cdot\textbf{E}_{s}^+(-\textbf{r})dS}{\int\hat{\textbf{e}}\cdot\boldsymbol{\mathcal{E}}{}^+(\textbf{r})dS},
\end{equation}
which, due to the symmetry, satisfy
\begin{equation}\label{Eq:tr}
t = 1+r.
\end{equation}
The (power) transmission and reflectivity of the lattice are defined by 
$T=|t|^2$ and $R=|r|^2$.
For a large subwavelength 2D array, only the zeroth order Bragg peak remains for the far-field light. The light is then scattered purely in the forward and backward direction,
and, for an excitation with a spatially uniform phase profile, 
the coherently transmitted light  at a point $(0,0,z)$ can be approximated by~\cite{dalibardexp,Javanainen17,Facchinetti18,Javanainen19}
\begin{equation}\label{EFieldSlab}
\begin{split}
\hat{\textbf{E}}{}^+(z)=\mathcal{E}_0\hat{\textbf{e}}e^{\text{i}kz} + \frac{\text{i}k\mathcal{D}}{2\mathcal{A}\epsilon_0}\sum_{l}^{}[\hat{\textbf{e}}-(\hat{\textbf{z}}\cdot\hat{\textbf{e}})\hat{\textbf{z}}]e^{\text{i}kz}\hat{\sigma}_l^-,
\end{split}
\end{equation}
when $\lambda \lesssim \xi \ll \sqrt{\mathcal{A}} $, where $\mathcal{A}$ is the total area of the array. 
Note that this zeroth-order diffraction is valid for any incident plane wave field, not just those normal to the lattice plane that we consider here.
Using Eqs.~\eqref{EFieldSlab} and~\eqref{Analytic}, we obtain 
for the $\textbf{q}=\textbf{0}$ mode,
\begin{equation}\label{Eq:Transmission2}
r = \text{i} (\gamma+\tilde{\gamma})\frac{\rho_{ge}}{\mathcal{R}}.
\end{equation}
Inserting the coherence from Eq.~\eqref{SScoherence} gives the reflection amplitude as a function of the excited level population. In fact, for a uniform excitation, Eq.~\eqref{EFieldSlab} coincides with the incident
and scattered light for a single-atom 
in 1D scalar electrodynamics~\cite{WaveguidePRA2017}, as the light is scattered only in the forward and backward directions, and the spatially uniform mode behaves as a single ``superatom''. In the LLI limit, $\rho_{ee}=0$ in Eq.~\eqref{SScoherence} and, at the resonance of the $\textbf{q}=\textbf{0}$ eigenmode ($\Delta=-\tilde\Omega$),
$\rho_{ge}=\text{ i} \mathcal{R}/(\gamma+\tilde\gamma) $, resulting in the total reflection $r=-1$, analogously to the total reflection of a resonant atom in 1D electrodynamics~\cite{Javanainen1999a}. 
The total resonance reflection from a planar array of linear dipole scatterers goes back to early electrodynamics~\cite{Tretyakov}, but has more recently been investigated in nanophotonics ~\cite{Laroche2006,Abajo07,CAIT}, with close to 100\% experimental realizations~\cite{Moitra2014,Moitra2015}, and has now similarly been highlighted for atoms~\cite{Bettles2016,Facchinetti16,Shahmoon}.

Using Eqs.~\eqref{SScoherence}, \eqref{Eq:Transmission2}, and~\eqref{Eq:tr}, the extinction $1-T$ and reflectivity $R$ can be obtained
\begin{subequations}\label{Eq:Reflectivity-2DUniformArray-Z}
\begin{align}
\begin{split}
1-T &= -\frac{Z(\gamma+\tilde{\gamma})\left[2(\gamma-Z\tilde{\gamma})+Z(\gamma+\tilde{\gamma})\right]}{(\Delta-Z\tilde{\Omega})^2+(\gamma-Z\tilde{\gamma})^2},
\end{split}\label{T(Z)}\\
\begin{split}
R &= \frac{Z^2(\gamma+\tilde{\gamma})^2}{(\Delta-Z\tilde{\Omega})^2+(\gamma-Z\tilde{\gamma})^2}.
\end{split}\label{R(Z)}
\end{align}
\end{subequations}
We also define the normalized flux through the array, 
\begin{equation}\label{normalised flux}
\begin{split}
F=\frac{ \int\langle \hat{\textbf{E}}{}^{-}(\textbf{r})\cdot\hat{\textbf{E}}{}^{+}(\textbf{r})\rangle dS}{\int |\boldsymbol{\mathcal{E}}{}^+(\textbf{r})|^2 dS},
\end{split}
\end{equation} 
where the incident flux is given by $\int |\boldsymbol{\mathcal{E}}{}^+(\textbf{r})|^2 dS = Na^2|\mathcal{E}_0|^2$ and the expectation value of the total field product can be expanded as
\begin{equation}\label{ElectricFieldExpansion}
\begin{split}
&\langle\hat{\textbf{E}}{}^-(\textbf{r})\cdot\hat{\textbf{E}}{}^+(\textbf{r})\rangle=|\boldsymbol{\mathcal{E}}{}^+(\textbf{r})|^2+\boldsymbol{\mathcal{E}}{}^-(\textbf{r})\cdot\langle\hat{\textbf{E}}{}_{s}^+(\textbf{r})\rangle+\\
&\langle\hat{\textbf{E}}{}_{s}^-(\textbf{r})\rangle\cdot\boldsymbol{\mathcal{E}}{}^+(\textbf{r})+|\langle\hat{\textbf{E}}{}_{s}^-(\textbf{r})\rangle|^2+\langle\delta\hat{\textbf{E}}{}_{s}^-(\textbf{r})\cdot\delta\hat{\textbf{E}}{}_{s}^+(\textbf{r})\rangle.
\end{split}
\end{equation}
The first term in Eq.~\eqref{ElectricFieldExpansion} contributes to the incident light intensity, while the next three terms are the coherently scattered light. The last term, $\langle\delta\hat{\textbf{E}}{}_{s}^-(\textbf{r})\cdot\delta\hat{\textbf{E}}{}_{s}^+(\textbf{r})\rangle =\langle\hat{\textbf{E}}{}_{s}^-(\textbf{r})\cdot\hat{\textbf{E}}{}_{s}^+(\textbf{r})\rangle-\langle\hat{\textbf{E}}{}_{s}^-(\textbf{r})\rangle\cdot\langle\hat{\textbf{E}}{}_{s}^+(\textbf{r})\rangle$, is due to incoherent scattering by position and quantum fluctuations. Because we consider atoms at fixed positions, it is here solely due to quantum fluctuations. 
The normalized flux for the incoherent light is given by
\begin{equation}\label{IncFluc1}
\begin{split}
F_{\rm{inc}}&=\frac{ \int\langle\delta\hat{\textbf{E}}{}_{s}^-(\textbf{r})\cdot\delta\hat{\textbf{E}}{}_{s}^+(\textbf{r})\rangle dS}{\int |\boldsymbol{\mathcal{E}}{}^+(\textbf{r})|^2 dS}\\
&=\sum_{l}^{N}\left(\rho_{ee}^{(l)}-|\rho_{ge}^{(l)}|^2\right)\frac{\int|\mathsf{G}(\textbf{r}-\textbf{r}_l)\textbf{d}_{ge} |^2dS}{\int |\boldsymbol{\mathcal{E}}{}^+(\textbf{r})|^2 dS},
\end{split}
\end{equation}
where all the quantum correlations between different atoms vanish in the last line due to the mean-field approximation we are using.
This expression differs from the usual semiclassical scattering description (which for the incoherent contribution would vanish for fixed atomic positions) due to the inclusion of the single-atom
quantum contribution from the $\rho^{(l)}_{ee}$ terms.
Collecting all the incoherently scattered light over a closed surface, the incoherent flux [substituting $\rho^{(l)}_{ee}=\rho_{ee}$ and $\rho_{ge}^{(l)}=\rho_{ge}$] is
\begin{equation}\label{IncFluc}
\begin{split}
F_{\rm{inc}}&=2\gamma(\gamma+\tilde{\gamma})\left(\frac{\rho_{ee}}{|\mathcal{R}|^2}-\left|\frac{\rho_{ge}}{\mathcal{R}}\right|^2\right)\\
&=2(\gamma+\tilde{\gamma})\text{Im}\left[\frac{\rho_{ge}}{\mathcal{R}}\right]-2(\gamma+\tilde{\gamma})^2\left|\frac{\rho_{ge}}{\mathcal{R}}\right|^2,
\end{split}
\end{equation}
where in the last line, we have used Eq.~\eqref{excitations} to eliminate the excited level population. 
Summing up the incident, coherent and incoherent normalized fluxes equals to one, as it should, implying that our model conserves energy. 
When driving a single spatially uniform mode in the large lattice limit where there is only the exact forward and backward scattering, and the total electric field is given by Eq.~\eqref{EFieldSlab}, the integration order in Eqs.~\eqref{Eq:TransmissionAmplitude} can be changed when calculating $R$ and $T$, with $|\int \hat{\textbf{e}}\cdot\textbf{E}^{+}(\textbf{r}) dS|^2 = \int |\textbf{E}^{+}(\textbf{r})|^2 dS$.
The normalized fluxes for the incident and coherently scattered light can then be replaced by $R$ and $T$ [Eqs.~\eqref{Eq:tr} and~\eqref{Eq:Transmission2}], and  we have
\begin{equation}\label{normalised flux2}
\begin{split}
R+T + F_{\rm{inc}} = 1,
\end{split}
\end{equation} 
which also follows from energy conservation, with $R+T<1$ indicating the presence of incoherent scattering.

\subsubsection{Transmission of cooperative and single-atom solutions}

In Fig.~\ref{Extinction}(a), we analyze the extinction of light as a function of intensity at $\Delta = 0$ for an array with lattice spacing $a=0.1\lambda$ and $\hat{\textbf{e}}=(1,1,0)/\sqrt{2}$, showing also the analytic approximate bistable solutions, Eqs.~\eqref{CoopandSingleAtomSoln}, which agree well with the numerics.
The inset of Fig.~\ref{Extinction}(a) shows an example of hysteresis by solving the dynamics of Eqs.~\eqref{SpinEqs} for the spatially uniform mode when the detuning is varied from an initial value of $\Delta/\gamma = -8$ ($\Delta/\gamma = 8$) to $\Delta = 0$, indicated by the right (left) arrows. 
We can see clearly the different behaviors of the extinction, which will result in observable changes in the light when the system jumps between the bistable solutions. 
The extinction of light from the cooperative solution is nearly constant, only changing by $\delta(1-T)\sim 10^{-2}$ from $I=0$ to $I/I_{\rm{sat}}=260$, as the atoms behave collectively and reflect the incident light.
Using the cooperative solution, Eq.~\eqref{CoopSoln}, to obtain $\rho_{ge}$ [Eq.~\eqref{Eq:Coherence}], an analytic low intensity expansion of the extinction and reflectivity in powers of $|\mathcal{R}|^2/|2C+1|^2$ is given by
\begin{subequations}\label{Eq:RCoop}
	\begin{align}
	\begin{split}
	&1-T\approx\frac{(\gamma+\tilde{\gamma})^2}{(\gamma+\tilde{\gamma})^2+(\Delta+\tilde{\Omega})^2}\left(1-\frac{4|\mathcal{R}|^4}{\eta^4}\frac{\text{Re}[2C+1]^2}{|2C+1|^8}\right),
	\end{split}\\
	\begin{split}
	&R\approx\frac{(\gamma+\tilde{\gamma})^2}{(\gamma+\tilde{\gamma})^2+(\Delta+\tilde{\Omega})^2}\left(1-\frac{4|\mathcal{R}|^2}{\eta^2}\frac{\text{Re}[2C+1]}{|2C+1|^4}\right).
	\end{split}
	\end{align}
\end{subequations}
For $|\mathcal{R}|^2=0$, the extinction and reflectivity are equal and both given by a Lorentzian centered on the collective line shift, $\Delta = -\tilde{\Omega}$, with maximum values of $1-T=1$ and $R=1$, respectively, and vanishing incoherent scattering, $F_{\rm{inc}}=0$.
As the intensity increases, the extinction and reflectivity begin to deviate. The extinction remains constant until the next leading order term, which is quadratic in intensity, while the reflectivity has a linear decrease with intensity as the excited level population increases and atoms start scattering light incoherently. Next order corrections scale inversely with the atomic density through $C$, highlighting how the cooperative nature of the atoms heavily suppress the transmitted and incoherently scattered light, and how even beyond the LLI, total extinction and reflection of light are possible.
Conversely, for the single-atom solution, the extinction and reflectivity both decrease with intensity as the atoms saturate and collective behavior is lost. Expanding the single-atom solution in powers of $\gamma^2/|\mathcal{R}|^2$ gives an approximate extinction and reflectivity at high intensities of
\begin{subequations}\label{Eq:RSA}
\begin{align}
\begin{split}
1-T&\approx\frac{(\gamma+\tilde{\gamma})}{|\mathcal{R}|^2}\left[\gamma+\frac{\eta^2\left(\tilde{\gamma}-\gamma\right)}{4|\mathcal{R}|^2}\right],
\end{split}\\
\begin{split}
R&\approx\frac{(\gamma+\tilde{\gamma})^2\eta^2}{4|\mathcal{R}|^4}.
\end{split}
\end{align}
\end{subequations}
The extinction decays inversely with the intensity, 
with the next leading order term recovering a quadratic dependence on the detuning. For large detunings, the single-atom solution no longer remains valid and previous studies in less densely packed arrays have shown how the lineshape will exhibit two symmetric peaks, analogously to the vacuum Rabi splitting in optical cavities~\cite{Bettles2020}. The reflectivity decays more rapidly with intensity than the extinction, indicating that some light is lost due to incoherent scattering.

\subsubsection{Maximum extinction}

\begin{figure}
	\hspace*{0cm}
	\includegraphics[width=\widthscale\columnwidth]{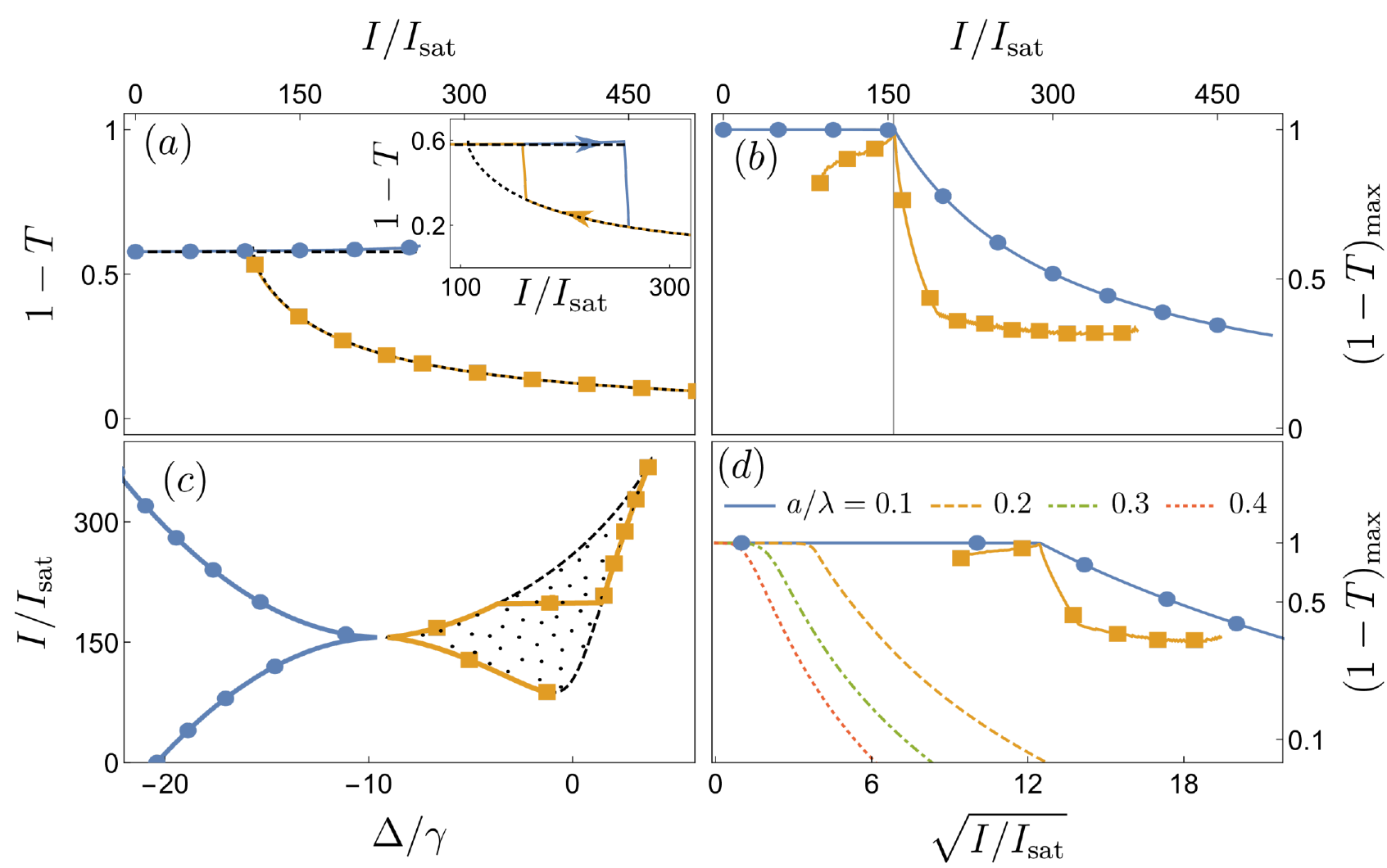}
	\vspace{-0.7cm}
	\caption{Extinction of light from a 2D planar array. 
	(a) Extinction of light, $1-T$, for an array with lattice spacing $a=0.1\lambda$ at $\Delta = 0$ for the cooperative (blue circles) and single-atom (orange squares) solutions with approximate cooperative (black dashed line) and single-atom (dot-dashed line) solutions also shown. 
	The inset shows hysteresis in the extinction from a negative (left arrow) and positive (right arrow) detuning sweep, resulting in a jump of the reflectivity.
	(b) Maximum extinction of light, $(1-T)_{\rm{max}}$, at any detuning. At the critical intensity (gray line), the extinction undergoes a sharp change in behavior. 
	(c) Detuning and intensity values of the extinction maximum for the solutions in (b). The maximum extinction of the single-atom solution lies along the edge of the bistability region (dotted region with black dashed line). The array completely extinguishes the incident light on resonance with the modified line shift, $\Delta = (2\rho_{ee}-1)\tilde{\Omega}$ below a critical intensity, $I_c/I_{\rm{sat}}\simeq155$. 
	(d) Maximum extinction for arrays with different lattice spacings. 
	}
	\vspace{0cm}
	\label{Extinction}
\end{figure}

In Fig.~\ref{Extinction}(b), we analyze the maximum extinction of light, $(1-T)_{\rm{max}}$, as a function of intensity, with the detuning of the maximum plotted in Fig.~\ref{Extinction}(c) (see Appendix~B for similar analysis of the reflectivity).
Two extinction solutions emerge, where below a critical intensity $I_c/I_{\rm{sat}}\simeq155$, the first extinction solution is nearly constant when the detuning is resonant with the modified collective line shift, $\Delta = (2\rho_{ee}-1)\tilde{\Omega}$, with $(1-T)_{\rm{max}}\simeq1$ at $I=0$ and $(1-T)_{\rm{max}}-1\sim 10^{-3}$ at $I_c$.
A second extinction solution appears when crossing the bistability region, with maximum values at $\Delta = (2\rho_{ee}-1)\tilde{\Omega}$, but with different $\rho_{ee}$. This solution initially extinguishes only $\sim80\%$ of the light, but counter-intuitively extinguishes more of the incident field as intensity increases, with $(1-T)_{\rm{max}}-1\sim 10^{-2}$ at $I_c$.
For $I>I_c$, the extinction of both solutions decreases and the lattice starts to become transparent.  

The near-complete extinction of the incident light and sharp change at the critical intensity are surprising, as $I_c$ is well beyond the LLI limit where we would expect the atoms to saturate and the extinction to smoothly decrease.
To explain this, we analyze the extinction, Eq.~\eqref{T(Z)}, which reaches maximum values for a given intensity when 
\begin{equation}\label{maxT}
\begin{split}
& \frac{d(1-T)}{d\Delta} = \frac{2(\gamma+\tilde{\gamma})(\Delta - Z\tilde{\Omega})f(Z)}{p'(Z)[(\Delta-Z\tilde{\Omega})^2+(\gamma-Z\tilde{\gamma})^2]^2}=0,
\end{split}
\end{equation}
where $p'(Z)$ is the derivative of Eq.~\eqref{SScubic} with respect to $Z$, and $f(Z)$ is a quartic in $Z$ (see Appendix~B). For $I<I_c$ ($I>I_c$), $\Delta = Z\tilde{\Omega}$ corresponds to an extinction maximum (minimum), while there is no minimum for $I<I_c$ and solutions to $f(Z)=0$ give the maximum for $I>I_c$.
Substituting $\Delta = Z\tilde{\Omega}$ into Eq.~\eqref{T(Z)} gives the maximum extinction for $I<I_c$,
\begin{equation}\label{Eq:Reflectivity-2DUniformArray-Z2}
\begin{split}
1-T &= -\frac{Z(\gamma+\tilde{\gamma})\left[2\gamma(1-Z)+Z(\gamma+\tilde{\gamma})\right]}{(\gamma-Z\tilde{\gamma})^2}.\\
\end{split}
\end{equation}
For a small excited level population, $(\gamma+\tilde{\gamma})Z\gg 2\gamma$ for closely spaced atoms where $(\gamma+\tilde{\gamma})\gg \gamma$, and Eq.~\eqref{Eq:Reflectivity-2DUniformArray-Z2} gives $1-T = 1$. For larger collective linewidths, the excited level population can take larger values with $(\gamma+\tilde{\gamma})Z\gg 2\gamma$ still being satisfied, and therefore $1-T\approx1$ holds for greater intensities.
The complete extinction of incident light can be understood by considering the collective uniform response of the array as a superatom. A large collective linewidth indicates that the atoms quickly re-emit the absorbed light to counteract the incident field, resulting in the array becoming highly reflective, and meaning the atoms have to be driven more strongly to saturate and for light to pass through.
Maximum extinction of the incident field is on resonance with the collective mode line shift where the atoms can absorb most strongly, with changes in the excited level population shifting the collective mode resonance. 

Increases in the excited level population eventually cause the collective behavior to break down at a critical intensity [found when the sign of $d^2(1-T)/d\Delta^2$ changes],
\begin{equation}\label{Eq:CriticalIntensity}
\begin{split}
\frac{I_c}{I_{\rm{sat}}} = \frac{(\gamma+\tilde{\gamma})^3(\tilde{\gamma}-2\gamma)^2}{4\gamma^2(\gamma-\tilde{\gamma})^2(\tilde{\gamma}-3\gamma)}\approx1+\frac{(\tilde{\gamma}+2\gamma)^2}{4\gamma^2}.
\end{split}
\end{equation}
In the last term we have expanded for $\tilde{\gamma} \gg \gamma$ and can see the critical intensity grows quadratically with the collective linewidth.
The excited level population at $I=I_c$ and $\Delta = Z\tilde{\Omega}$ is
\begin{equation}
\rho_{ee} = \frac{1}{4}\frac{(\gamma+\tilde{\gamma})}{(\tilde{\gamma}-\gamma)} \approx \frac{1}{4},
\end{equation}
which is halfway towards the atoms being completely saturated. 
For $I>I_c$, $\rho_{ee}$ gradually increases from $\rho_{ee}\approx 1/4$ to $\rho_{ee}\approx1/2$ for the extinction maximum when $f(Z)=0$, but rapidly increases to $\rho_{ee}\approx 1/2$ for $\Delta = Z\tilde{\Omega}$. The extinction then becomes a minimum as the atoms have saturated and cannot absorb anymore of the incident light. Spontaneous emission from incoherent scattering also reaches maximum values, with $F_{\rm{inc}}$ dropping to zero and therefore most light passes through the array. 

The sharp change in the extinction can only occur for sufficiently closely-packed lattices, with no positive solution to Eq.~\eqref{Eq:CriticalIntensity} when $\tilde{\gamma}\leq3\gamma$.
Using Eq.~\eqref{Analytic}, the critical lattice spacing is $a_c \simeq 0.244\lambda$. For $a>a_c$, $\Delta = Z\tilde{\Omega}$ always gives the maximum extinction, which smoothly decreases with intensity, and can be seen in Fig.~\ref{Extinction}(d), with the region of $1-T\approx 1$ moving to smaller intensities as the lattice spacing is increased.
Eventually, for larger atom spacings, the strong superradiant contribution to the extinction is lost and the array begins to become transparent even at low intensities~\cite{Bettles2020}.
For small lattice spacings where bistability emerges, the maximum extinction at $I=I_c$ can be found to coincide with the tri-critical point of the bistability region, as seen in Fig.~\ref{Extinction}(c).
Finding when the discriminant of Eq.~\eqref{SScubic} is zero for $\Delta = Z\tilde{\Omega}$ gives intensity bounds for the bistability region of
\begin{equation}\label{Eq:CriticalIntensityBistability}
\begin{split}
\frac{I}{I_{\rm sat}} =\frac{\tilde{\gamma}^2+20\gamma\tilde{\gamma}-8\gamma^2}{8\gamma^2}\pm\frac{\sqrt{(\tilde{\gamma}-8\gamma)^3\tilde{\gamma}}}{8\gamma^2},
\end{split}
\end{equation}
where the threshold $\tilde{\gamma}>8\gamma$ from Eq.~\eqref{thresh} naturally arises for the intensity to be real.
In the limit $\tilde{\gamma} \gg \gamma$, the upper bound of Eq.~\eqref{Eq:CriticalIntensityBistability} is approximately $I/I_{\rm{sat}} \approx 1+(\tilde{\gamma}+2\gamma)^2/4\gamma^2$, which agrees with Eq.~\eqref{Eq:CriticalIntensity}, and so the maximum extinction changes when reaching the tri-critical point of the bistability region.

\subsection{Group Delay}

Transmission of light through a lattice also results in a phase shift,  which can be quantified by the group delay
\begin{equation}\label{Eq:GroupVelocity}
	\tau_g(\Delta) = \frac{d\arg[t(\Delta)]}{d\Delta}.
\end{equation}
Strong group delays represent significant phase shifts in the incident light, leading to large delays in the amplitude envelope of a pulse traveling through the lattice. 
Collective interactions in arrays have been shown to lead to particularly large group delays in the LLI limit when coupled to narrow subradiant LLI eigenmodes~\cite{CAIT,Facchinetti16}, with possible applications in enhanced sensing.

\subsubsection{Group delay of cooperative and single-atom solutions}
\label{groupdelaysolns}

\begin{figure}
	\hspace*{0cm}
	\includegraphics[width=\columnwidth]{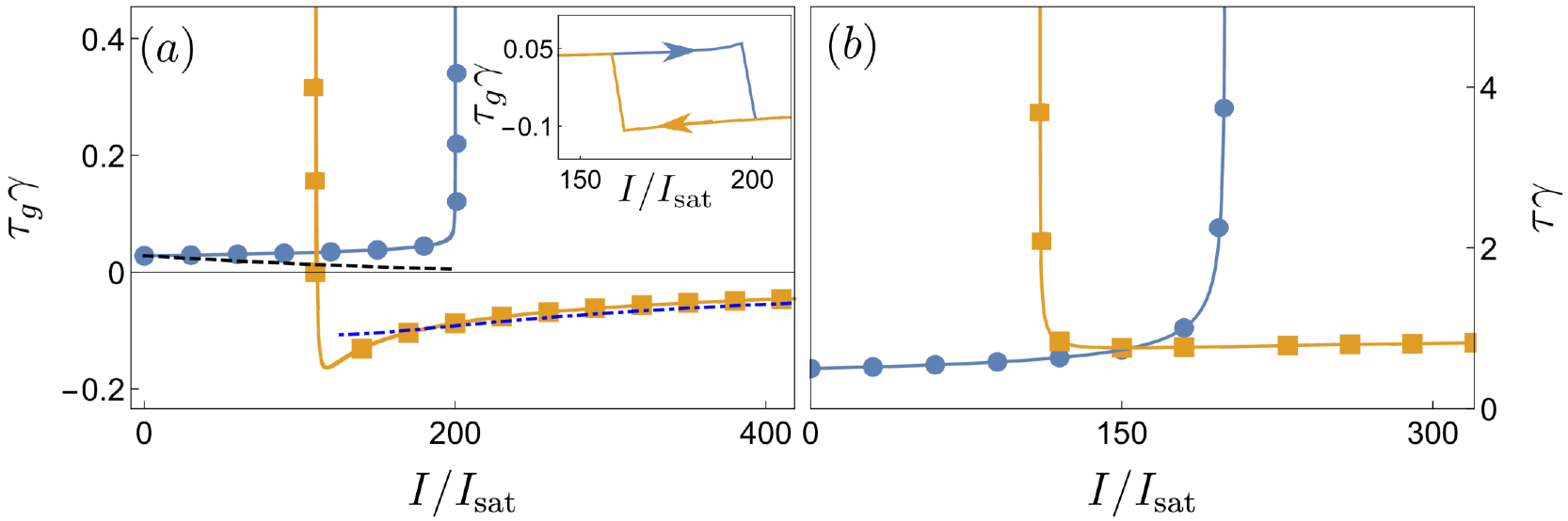}
	\vspace{-0.7cm} 
	\caption{
		Group delay and critical slowing of a pulse through an array of atoms with lattice spacing $a=0.1\lambda$, $\hat{\textbf{e}} = (1,0,0)$ and $\Delta/\gamma = -3.5$. 
		(a) Group delay of the spatially uniform mode for the cooperative (blue circles) and single-atom (orange squares) solutions with analytic estimates Eq.~\eqref{Eq:LLIgroupdelay} (black-dashed) and Eq.~\eqref{Eq:SAgroupdelay} (blue-dot-dashed). The group delay divergences at the bistability region edge. 
		The inset shows hysteresis in the group delay from a negative (left arrow) and positive (right arrow) detuning sweep.
		(b) Longest decay time, $\tau=-1/\text{Re}[\lambda]$, for the dynamics to settle to the steady state for the cooperative and single-atom solutions. Critical slowing occurs at the edge of the bistability region, with a diverging decay time.
	}
	\vspace{0cm}
	\label{Fig:GroupDelay}
\end{figure}

Figure~\ref{Fig:GroupDelay}(a) shows an example of the group delay for both bistable solutions at $\Delta/\gamma=-3.5$,
with the inset showing hysteresis by evolving Eqs.~\eqref{SpinEqs} in time for the spatially uniform phase,
varying the detuning from an initial value of $\Delta/\gamma = -8$ ($\Delta/\gamma = 8$) to $\Delta/\gamma=-3.5$, indicated by the right (left) arrows.
We find the cooperative and single-atom solutions give positive and negative group delays, respectively, which will result in a sharp phase change in light transmitted through the lattice when the system jumps between the bistable solutions.
For low intensities, where $\mathcal{R}_{\rm{eff}}\approx \mathcal{R}/(2C+1)$, the group delay is approximately
\begin{equation}\label{Eq:LLIgroupdelay}
\begin{split}
&\tau_g \approx \frac{(\gamma+\tilde{\gamma})\left[(\Delta+\tilde{\Omega})^2-2|\mathcal{R}|^2\right]}{4|\mathcal{R}|^4+(\Delta+\tilde{\Omega})^2\left[4|\mathcal{R}|^2+(\gamma+\tilde{\gamma})^2+(\Delta+\tilde{\Omega})^2\right]},
\end{split}
\end{equation}
while at large intensities where $\mathcal{R}_{\rm eff}\approx\mathcal{R}$, the group delay is approximately
\begin{equation}\label{Eq:SAgroupdelay}
\begin{split}
&\tau_g\approx\frac{-(\gamma+\tilde{\gamma})[2|\mathcal{R}|^2+\gamma(\gamma-\tilde{\gamma})-\eta^2]}{4|\mathcal{R}|^4+4|\mathcal{R}|^2[\eta^2-\gamma(\gamma+\tilde{\gamma})]+\eta^2(\Delta^2+\tilde{\gamma}^2)},
\end{split}
\end{equation}
which vanishes as $|\mathcal{R}|^2\rightarrow0$ as the atoms saturate and the array becomes transparent.

At the upper and lower boundaries of the bistability region, the group delay diverges for the vanishing solution.
The sign of the divergence changes with detuning, and is negative (positive) for the cooperative (single-atom) solution when $-1.62\alt\Delta/\gamma\alt-0.99$, and the divergences for both solutions are negative (positive) when $\Delta/\gamma \agt -0.99$ ($\Delta/\gamma \alt -1.62$).
Small shifts in the intensity and detuning lead to large changes in the group delay in the vicinity of the divergence, e.g., changing  $\Delta/\gamma =-3.5$ to $\Delta/\gamma =-3.499$ results in a drop of $\tau_g\gamma \simeq 1.34$ to $\tau_g\gamma \simeq 0.3$ in the vicinity of the divergence at $I/I_{\rm{sat}}\simeq200$ in Fig.~\ref{Fig:GroupDelay}(a).
As a function of $Z$, the group delay is given by
\begin{equation}\label{Eq:GroupDelay}
\begin{split}
&\tau_g = \frac{(\gamma+\tilde{\gamma})Q(Z)}{[\Delta'^2+\gamma'^2][\Delta'^2+\gamma^2(1+Z)^2]p'(Z)},
\end{split}
\end{equation}
where
\begin{equation}\label{Eq:GroupDelayextra}
\begin{split}
&\gamma' = \gamma - Z\tilde{\gamma}, \quad \Delta' = \Delta - Z\tilde{\Delta},
\end{split}
\end{equation}
and
\begin{equation}
\begin{split}
&Q(Z)=Zp'(Z)[\Delta'^2-\gamma\gamma'(1+Z)]-2(1+Z)\Delta'\times\\
&\phantom{Q(Z)=}[\gamma\gamma'\Delta'+(\gamma^2\Delta'-\gamma\gamma'\Delta)(1+Z)+\Delta\Delta'^2].
\end{split}
\end{equation}
Crucially, Eq.~\eqref{Eq:GroupDelay} depends on the derivative of Eq.~\eqref{SScubic} with respect to $Z$, $p'(Z)$, which goes to zero at the bistability boundary and results in the group delay divergence.
\subsubsection{Critical slowing}

Bistability is associated with the presence of a first-order phase transition and critical slowing, where increasingly longer times are needed to reach the steady state~\cite{Bonifacio1978,Carr2013,Bonifacio1979,Scheffer2009} at the edge of the bistability region. 
Critical slowing can be shown by linearizing Eqs.~\eqref{SpinEqs} about the spatially uniform stationary state, resulting in a matrix equation $\delta \dot{\boldsymbol{\rho}} = \textbf{M}\delta \boldsymbol{\rho}$ where $\delta \boldsymbol{\rho}= (\delta S_x,\delta S_y,\delta \rho_{ee})$, with $S_x+\text{i}S_y=\rho_{ge}$ and
\begin{equation}
\begin{split}
\textbf{M}=\begin{pmatrix}
-\gamma' & -\Delta' & 2S_x\tilde{\gamma}+2S_y\tilde{\Omega} \\
\Delta' & -\gamma' & 2S_y\tilde{\gamma}-2S_x\tilde{\Omega}-2\mathcal{R}\\
-4S_x\tilde{\gamma} & -4S_y\tilde{\gamma}+2\mathcal{R} & -2\gamma
\end{pmatrix}.
\end{split}
\end{equation}
The eigenvalues of this matrix obey
\begin{equation}\label{Eq:CharPol}
\begin{split}
\lambda^3+a_2\lambda^2+a_1\lambda+a_0=0,
\end{split}
\end{equation}
with coefficients [simplified using Eqs.~\eqref{SScoherence} and~\eqref{SScubic}] 
\begin{equation}
\begin{split}
&a_2=2\gamma+2\gamma',\\
&a_1=\frac{\gamma'^2+\Delta'^2+(Z-1)\gamma\gamma'-(Z+1)(2\gamma^2+\Delta \Delta')}{Z^2},\\
&a_0=-\frac{2\gamma p'(Z)}{Z}.\\
\end{split}
\end{equation}
The sign of the real part of the eigenvalues determines whether fluctuations about a steady-state decay ($\text{Re}[\lambda]<0$) or grow ($\text{Re}[\lambda]>0$), indicating instability. The last coefficient, $a_0$, vanishes at the edge of the bistability region where $p'(Z) =0$, which also results in the group delay divergence  (Sec.~\ref{groupdelaysolns}). With $a_0=0$, $\lambda=0$ is an eigenvalue, leading to critical slowing where small fluctuations from the steady state take an infinitely long time to relax, with a decay time of $\tau = -1/\text{Re}[\lambda]$, as illustrated in Fig.~\ref{Fig:GroupDelay}(b).
The decay time is sensitive to small changes in intensity and detuning in the vicinity of the divergence, e.g., for $I/I_{\rm{sat}}=200$ in Fig.~\ref{Fig:GroupDelay}(b), changing $\Delta/\gamma =-3.5$ to $\Delta/\gamma =-3.499$ results in a 5-fold decrease of $\tau$ from $\tau\gamma \simeq 151$ to $\tau\gamma \simeq30$.
This significant drop in the decay times means only a small parameter change is needed to avoid critical slowing, which therefore makes measuring observables such as the group delay experimentally easier as the system no longer takes a long time to relax to the steady state. However, as discussed in Sec.~\ref{groupdelaysolns}, changing $\Delta/\gamma =-3.5$ to $\Delta/\gamma =-3.499$ also results in a near 5-fold decrease in the group delay, with $\tau_g\gamma = 1.34$ dropping to $\tau_g\gamma = 0.3$, and therefore it may be challenging to avoid critical slowing and obtain large group delay values simultaneously.

\section{Concluding remarks}\label{discussion}

We analyzed the mean-field behavior of a closely packed array of atoms, where intrinsic optical bistability emerges between steady-state solutions with different spatially uniform excited level populations. 
We developed a theory for optical bistability that provides threshold conditions, in some cases even analytically, for the lattice spacings, intensities and detunings needed for bistability to emerge. 
Our theory bears strong similarities to the theory of bistability in optical cavity systems, where for a specific detuning, a direct analogy between cooperativity due to the cavity mirrors and cooperativity due to
the radiative long-range DD interactions in a free-space atomic arrays is established.
While the emphasis was on uniform systems, we also found that bistability depends on the collective linewidths of the underlying LLI eigenmodes of the excitations, with interesting possibilities even to
study bistabilities between a superradiant and subradiant mode~\cite{parmee2020}. The edges of the bistability regions are associated with phase transitions, critical slowing down, and large group delays.
Moreover, we showed that driving the spatially uniform LLI eigenmode leads to the array completely extinguishing the incident field up to a critical intensity, $I_c/I_{\rm{sat}} \simeq 155$, which extends  LLI results~\cite{Tretyakov,Laroche2006,Abajo07,CAIT,Moitra2014,Moitra2015,Bettles2016,Facchinetti16,Shahmoon,Rui2020} to the nonlinear regime at much higher intensities.

In our model we solved the nonlinear optical response by including the full internal level dynamics of each individual atom and the scattering processes between the atoms, based on their discrete spatial positions, but ignored light-induced quantum correlations between the different atoms. Although we considered two-level atoms at fixed lattice sites, the general semiclassical theory~\cite{Lee16} incorporates the full multilevel structure and fluctuations of the atomic positions, and it has been applied, e.g., to the simulations of optical pumping between different electronic ground levels in a trapped dense atomic ensemble~\cite{Machluf2018}. An obvious advantage is that the number of equations scales linearly with the atom number $N$, while the size of the density matrix for the full quantum solution $\sim 2^{2N}$ quickly becomes intractable for larger systems.
There is, however, also a more fundamental difference: the evolution of the full quantum many-body density matrix is linear and cannot exhibit nonlinear bistable behavior that is inherently a classical phenomenon. Classical nonlinear phenomena, such as bistabilities and dynamical instabilities, emerge from a quantum system due to decoherence~\cite{Joos85,Zurek03,Walls85} or continuous quantum measurement-induced back-action~\cite{Javanainen97,Javanainen_2013}. While in pristine experimental conditions quantum entanglement between the atoms could be preserved, noise, e.g., from magnetic fields or continuous monitoring of scattered light could quickly drive the system to the classical mean-field regime to display bistability. Controlling the experimental noise or changing the measurement scheme in such systems could potentially even be utilized for investigating quantum-classical interface and decoherence, and the emergence of classical nonlinear dynamics from a quantum system.

\section{Acknowledgments}
We acknowledge financial support from the UK EPSRC (Grant Nos.\ EP/S002952/1, EP/P026133/1)

\begin{appendices}
%%%%%%%%%%%%%%%%%%%%%%%%%%%%%%%%%%%%%%%
%

%%%%%%%%%%%%%%%%%%%%%%%%%%%%%%%%%%%%%%%%%%%%%%%%%%%%%%%%%%%%
\section{Comparison with cavities}
We consider $N$ cooperatively coupled two-level atoms in a single-mode cavity experiencing the same field~\cite{Tavis1968a}, with the master equation
\begin{equation}
\begin{split}
\dot{\hat{\rho}} = &-\frac{\text{i}}{\hbar}[\hat{H},\hat{\rho}] + \kappa(2\hat{a}\hat{\rho} \hat{a}^{\dagger}-\hat{a}^{\dagger}\hat{a}\hat{\rho}-\hat{\rho} \hat{a}\hat{a}^{\dagger}) \\
&+N\gamma(2\hat{\sigma}^{-}\hat{\rho}\hat{\sigma}^{+}-\hat{\sigma}^{-}\hat{\sigma}^{+}\hat{\rho}-\hat{\rho}\hat{\sigma}^{-}\hat{\sigma}^{+}),
\end{split}
\end{equation}
where $\hat{a}$ is a photon annihilation operator for the cavity mode, $\kappa$ the cavity linewidth, and $\gamma$ the free-space linewidth.
The Hamiltonian is
\begin{equation}
\begin{split}
&\hat{H} = \hbar\Delta_c\hat{a}^{\dagger}\hat{a}+\hbar\zeta(\hat{a}^{\dagger}+\hat{a})+\hbar\Delta N\hat{\sigma}^{+}\hat{\sigma}^{-}\\
&\phantom{\hat{H}}+\hbar gN(\hat{\sigma}^{-}\hat{a}^{\dagger}+\hat{\sigma}^{+}\hat{a})+\hbar N(\mathcal{R}\hat{\sigma}^{+}+\mathcal{R}^*\hat{\sigma}^{-}),
\end{split}
\end{equation}
where $\Delta_c$ ($\Delta$) is the detuning of the cavity field (atom) from the laser frequency, both of which we now set to zero, $g$ is the cavity-atom coupling, $\zeta$ is the field driving the cavity mode and $\mathcal{R}$ represents coherent transverse field driving the atoms.
The equations of motion with the decorrelation approximation $\langle \hat{a} \hat{\sigma}{}^{-} \rangle \approx \langle \hat{a} \rangle \langle \hat{\sigma}{}^{-} \rangle$ yield  [$\langle\hat{\sigma}^z\rangle=2\rho_{ee}-1$, $\langle\hat{\sigma}^{-}\rangle = \rho_{ge}$]
\begin{subequations}\label{cavityeoms}
	\begin{align}
	&\dot{\rho}_{ee}=-2\gamma\rho_{ee}+2\text{Im}[(\mathcal{R}^*+g\langle\hat{a}^{\dagger}\rangle) \rho_{ge}],\\
	&\dot{\rho}_{ge}=\left(\text{i}\Delta-\gamma\right)\rho_{ge}-\text{i}(2\rho_{ee}-1)(\mathcal{R}+g\langle\hat{a}\rangle),\\
	&\langle\dot{\hat{a}}\rangle=-\kappa\langle\hat{a}\rangle+\text{i}Ng\rho_{ge}+\text{i}\zeta.\label{aeq}
	\end{align}
\end{subequations}
Comparison with Eqs.~\eqref{SpinEqsMacro3} allows us to identify a new effective field
\begin{equation}
\begin{split}
\mathcal{R}'_{\rm{eff}}=\mathcal{R}+g\langle\hat{a}\rangle.
\end{split}
\end{equation}
The solutions of $\rho_{ee}$ and $\rho_{ge}$ from Eqs.~\eqref{cavityeoms} in terms of $\mathcal{R}'_{\rm{eff}}$ are the same as Eqs.~\eqref{Eq:EOMSolns}. From  Eq.~\eqref{aeq} we obtain a relationship between the input field and effective field
\begin{equation}
\begin{split}
&\mathcal{R}'_{\rm{eff}}\left[-\kappa-\frac{Ng^2\gamma}{\gamma^2+2|\mathcal{R}'_{\rm{eff}}|^2}\right]=-\text{i}g\zeta-\kappa\mathcal{R}.
\end{split}
\end{equation}
This can be simplified to the relationship between the incident and internal field in a cavity
\begin{equation}\label{cavityrel}
\begin{split}
y=x\left(1+\frac{2C}{1+|x|^2}\right),
\end{split}
\end{equation}
where we have defined the variables
\begin{equation}
\begin{split}
&x=\frac{\sqrt{2}\mathcal{R}'_{\rm{eff}}}{\gamma},\quad y=\frac{\sqrt{2}\mathcal{R}}{\gamma}-\frac{\text{i}g}{\kappa}\frac{\sqrt{2}\zeta}{\gamma},
\end{split}
\end{equation}  
and introduced the cooperativity parameter
\begin{equation}\label{generalC}
\begin{split}
C=\frac{Ng^2}{2\kappa\gamma},
\end{split}
\end{equation}
which is the standard cooperativity parameter for a cavity~\cite{Bonifacio1978,CarmichaelVol2}.
Equation~\eqref{Eq:Modyequation} in the main text can also exactly be cast in the form of Eq.~\eqref{cavityrel} when $\Delta/\gamma = \tilde{\Omega}(\textbf{q})/\tilde{\gamma}(\textbf{q})$ by substituting
\begin{equation}
\begin{split}
&\mathcal{R}\rightarrow y\frac{\gamma}{\tilde{\gamma}(\textbf{q})}\sqrt{\frac{\tilde{\gamma}(\textbf{q})^2+\tilde{\Omega}(\textbf{q})^2}{2}},\\
&\mathcal{R}_{\rm{eff}} \rightarrow x\frac{\gamma}{\tilde{\gamma}(\textbf{q})}\sqrt{\frac{\tilde{\gamma}(\textbf{q})^2+\tilde{\Omega}(\textbf{q})^2}{2}},
\end{split}
\end{equation}
and the cooperativity parameter $C= \tilde{\gamma}(\textbf{q})/2\gamma$.
In the limit that $\kappa = 0$, we can instead obtain for the cavity
\begin{equation}
y' = \frac{2C'x}{1+|x|^2},
\end{equation}
with $y'=-\text{i}\sqrt{2}\zeta/\gamma$, with a new cooperativity parameter $C'=Ng/2\gamma$. By comparison with the array, we find that $\tilde{\gamma}(\textbf{q})$ plays a similar role to the atom-cavity coupling $Ng$.

%%%%%%%%%%%%%%%%%%%%%%%%%%%%%%%%%%%%%%%%%%%%%%%%%%%%%%%%%
\section{Extinction and reflectivity}

\begin{figure}
	\hspace*{0cm}
	\includegraphics[width=\widthscale\columnwidth]{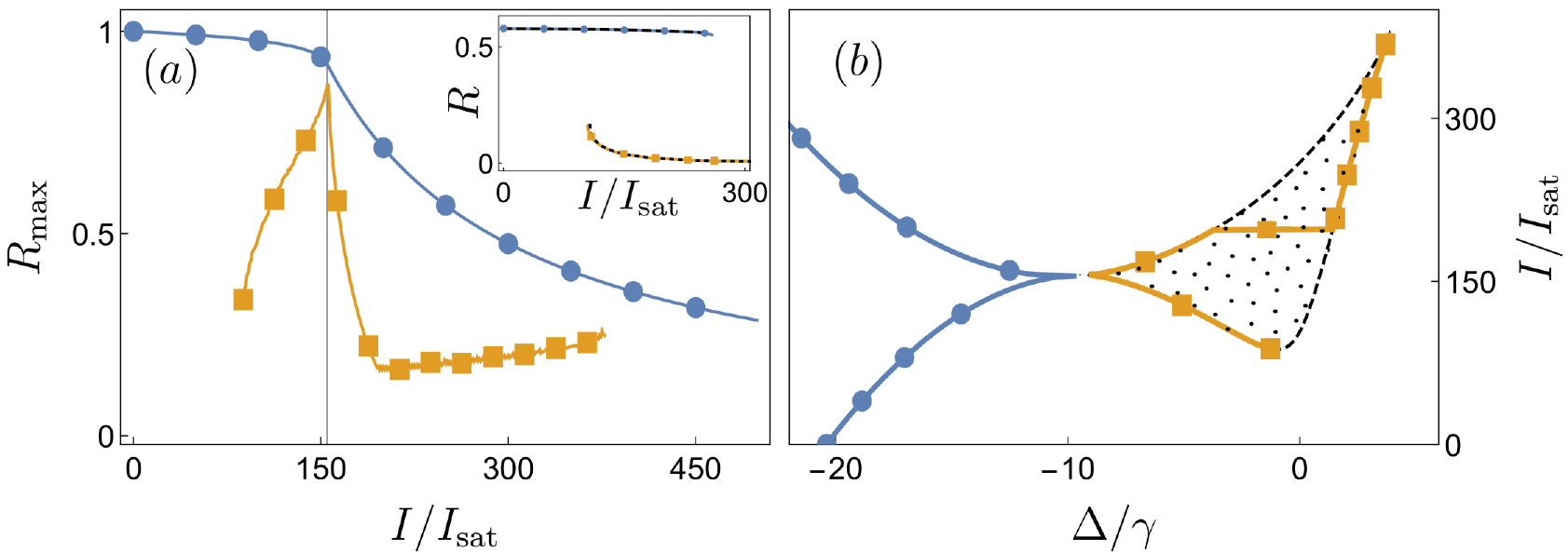}
	\caption{Reflectivity of a 2D planar array. 
		(a) Maximum reflectivity, $R_{\rm{max}}$, at any detuning for the cooperative (blue circles) and single-atom (orange squares) solutions. At the critical intensity (gray line), the reflectivity undergoes a sharp change in behavior. 
		The inset shows the reflectivity, $R$, at $\Delta = 0$ and approximate cooperative (black dashed line) and single-atom (dot-dashed line) solutions. 
		(b) Detuning and intensity values of the reflectivity maximum for the two solutions in (a). The maximum reflectivity of the single-atom solution lies along the edge of the bistability region (dotted region with black dashed line). The array reflects a significant portion of the incident light on resonance with the modified line shift, $\Delta = (2\rho_{ee}-1)\tilde{\Omega}$ below a critical intensity, $I^R_c/I_{\rm{sat}}\simeq154$. 
	}
	\vspace{0cm}
	\label{Reflectivity}
\end{figure}

The maximum of the extinction is found when Eq.~\eqref{maxT} is obeyed.
For $I<I_c$ ($I>I_c$), $\Delta = Z\tilde{\Omega}$ gives the maximum (minimum) extinction, while $f(Z)=0$ gives the maximum for $I>I_c$, where
\begin{equation}\label{Eq:f(Z)}
\begin{split}
&f(Z) =  2|\mathcal{R}|^2Z\left[3\gamma-Z\gamma-\gamma'\right]+\\
&\left[(2+3Z)\gamma'+3\gamma Z(1+Z)\right][\Delta'^2+\gamma'^2],
\end{split}
\end{equation}
where $\Delta'$ and $\gamma'$ are given by Eq.~\eqref{Eq:GroupDelayextra}.

Figure~\ref{Reflectivity}(a) shows the maximum reflectivity as a function of intensity, with the corresponding detuning in Fig.~\ref{Reflectivity}(b). 
The maximum reflectivity is on resonance with the collective lineshift below a critical intensity, $I_c^R$, decreasing from $R\simeq1$ at $I=0$ to $R\simeq0.939$ at $I_c^R/I_{\rm{sat}}\simeq154$.
The inset in Fig.~\ref{Reflectivity}(a) shows the reflectivity at $\Delta=0$. Similar to the extinction in Fig.~\ref{Extinction}(a), at low intensities the reflectivity is nearly constant, decreasing slightly with intensity as shown in Eq.~\eqref{Eq:RCoop}, with a change of $\delta R \sim 10^{-2}$ from $I=0$ to $I/I_{\rm{sat}}=260$, while decaying very quickly at high intensities due to the $1/|\mathcal{R}|^4$ term seen in Eq.~\eqref{Eq:RSA}.
The maximum reflectivity is found when 
\begin{equation}
\begin{split}
\frac{dR}{d\Delta} = -2(\gamma+\tilde{\gamma})^2\Delta'\frac{Zh(Z)}{p'(Z)[\gamma'^2+\Delta'^2]^2}  = 0,
\end{split}
\end{equation}
where $h(Z)$ is a cubic,
\begin{equation}\label{Eq:h(Z)}
\begin{split}
h(Z)=&2|\mathcal{R}|^2Z+(2+3Z)[\gamma'^2+\Delta'^2].
\end{split}
\end{equation}
For $I<I_c^R$ ($I>I_c^R$), $\Delta = Z\tilde{\Omega}$ is a minimum (maximum), while solutions to $h(Z)=0$ give the maximum for $I>I_c^R$.
The critical intensity for the reflectivity, found when the sign of $d^2R/d\Delta^2$ changes, is
\begin{equation}\label{Eq:CriticalIntensityR}
\begin{split}
&\frac{I_c^R}{I_{\rm{sat}}} =\frac{(\tilde{\gamma}+2\gamma)^2}{4\gamma^2},
\end{split}
\end{equation}
which differs from Eq.~\eqref{Eq:CriticalIntensity}, but with both values similar for closely-spaced arrays. Substituting $\Delta = Z\tilde{\Omega}$ into Eq.~\eqref{Eq:Reflectivity-2DUniformArray-Z} gives the maximum reflectivity for $I<I_c^R$,
\begin{equation}\label{Eq:Ronres}
\begin{split}
R &= \frac{Z^2(\gamma+\tilde{\gamma})^2}{(\gamma-Z\tilde{\gamma})^2}.
\end{split}
\end{equation}
For small excited level populations, $R\approx 1$, provided $\tilde{\gamma}\gg \gamma$ and $\gamma-Z\tilde{\gamma} \approx Z\tilde{\gamma}$.
 
\end{appendices}

%\bibliography{atomlight}


\end{document}